\documentclass[11pt,a4paper]{article}

\usepackage{bm, amssymb, amsmath, graphicx, mathtools, float}
\usepackage{lineno,hyperref,color}
\usepackage{graphicx}
\usepackage{fancyhdr}
\usepackage{longtable,tabularx}
\usepackage{multirow}
\usepackage{subfig}
\usepackage[backend=bibtex,sorting=none,maxbibnames=99,firstinits=true]{biblatex}

\fancypagestyle{firststyle}
{
	\fancyhf{}
	\fancyfoot[RE,LO]{}
	\fancyfoot[LE,RO]{Page \thepage}
}

\begin{document}
	
	\title{A set of orbital elements to fully represent the zonal harmonics around an oblate celestial body}
	
	\author{David Arnas\thanks{Massachusetts Institute of Technology, MA, USA. Email: \textsc{arnas@mit.edu}}, Richard Linares\thanks{Massachusetts Institute of Technology, MA, USA. Email: \textsc{linaresr@mit.edu}}}
	
	\date{}	
	
	\maketitle{} 	
	
	\thispagestyle{firststyle}
	
\begin{abstract}
This work introduces a new set of orbital elements to fully represent the zonal harmonics problem around an oblate celestial body. This new set of orbital elements allows to obtain a complete linear system for the unperturbed problem and, in addition, a complete polynomial system when considering the perturbation produced by the zonal harmonics from the gravitational force of an oblate celestial body. These orbital elements present no singularities and are able to represent any kind of orbit, including elliptic, parabolic and hyperbolic orbits. In addition, an application to this formulation of the Poincar\'e-Lindstedt perturbation method is included to obtain an approximate first order solution of the problem for the case of the $J_2$ perturbation.
\end{abstract}



\section{Introduction}

In celestial mechanics, analytical solutions and methods of analysis to study the motion in a gravitational field are of extreme importance for both scientific and engineering applications. These techniques allow, for instance, to generate new tools for the fast long-term propagation of orbiting objects, to analyze and better understand the dynamics of these systems, to evaluate the long term behavior of space debris (\cite{casanova2015long}), to study the stability of natural orbits (\cite{broucke1969stability}), or to define and assess satellite constellations (\cite{mcgrath2019general}), among other topics. To that end, a large variety of different formulations have been investigated over the years to study the long term evolution of the so-called main satellite problem, that is, the description of the dynamic of a orbiting object subjected to the perturbation produced by the oblateness of the central body (e.g. the Earth). Early examples of that are the solutions proposed by \cite{brouwer}, \cite{kozai1959motion}, \cite{deprit1969}, \cite{liu1974satellite}, which have been used extensively both in theoretical and applied problems.

The main satellite problem, even for the case of just considering the perturbation produced by the oblateness of a celestial body ($J_2$ term from the gravitational potential) has no analytical solution (\cite{irigoyen1993non,celletti1995non}). Therefore, several sets of orbital elements have been proposed over the years to study this problem and to ease the generation of approximate solutions. For instance, the classical or Kepler elements are the most widespread orbital elements in astrodynamics due to their simplicity and geometrical representation. In particular, classical variables define the state of a orbiting object based on the semi-major axis, the eccentricity, the inclination, the argument of perigee, the right ascension of the ascending node, and the true anomaly of its orbit. However, and when studying orbital perturbations, it is more convenient to use of the action-angle variables associated with the dynamics, which, for the case of the perturbed two-body problem, are the Delaunay variables (\cite{deprit1970main,barrio2003high,lara2014delaunay,lara2020exploring,lara2020solution}). Another well used set of orbital elements are the nodal variables, also known as Whittaker or Hill variables (\cite{hill1913motion,aksnes1972use,cid1969perturbaciones,cid1971perturbaciones}). This is a set of canonical variables that was devised to address some of the problems of definition that Delaunay variables present at particular orbits, for instance, near circular orbits. Also, and with a similar objective, \cite{broucke1972equinoctial} proposed the equinoctial variables, a set of elements that present no singularities in their formulation, which has been been applied in analytical, semi-analytical and numerical applications. Another non-singular set of variables used to deal with this problem are spherical coordinates (\cite{vinti1960theory}). This trend of defining new sets of variables to ameliorate the complexities in analytical satellite theories has continued in the literature (\cite{lara2019new,abad2001short,abad2020integration}), and is the aim of this work to continue that development by proposing a new set of variables that have a unique set of properties and that allows the application of relativity simple perturbation theory methods. 

All these sets of orbital elements were devised to be used alongside different perturbation theories in order to generate approximate analytical solutions to the zonal harmonics problem. To that end, and in a seminal work, \cite{brouwer} proposed a first order solution based on the von Zeipel method that later was complemented by \cite{lyddane1963small} and \cite{cohen1981radius} to account for small eccentricities and inclinations, and by \cite{coffey1986critical} to study orbits close to the critical inclination. \cite{kozai1959motion}, on the other hand, presented an alternative approach to generate a first order solution based on a decomposition of the dynamics in first-order secular, second-order secular, short-periodic, and long-periodic terms, and later, he proposed a second order solution \cite{kozai1962second} to the main satellite problem to extend Brouwer's formulation. 

Some years later, \cite{deprit1969} developed the use of Lie series to define a set of canonical mappings in the form of power series in a small parameter. This methodology allows to obtain the approximate solution to the system by an iterative recursive transformation (\cite{kamel1969expansion}) that can be extended to an arbitrary order of the solution. This perturbation technique has lead to the generation of the so called Lie-Deprit methods, which have been used extensively in the literature (\cite{kamel1969expansion,kamel1970perturbation,deprit1970main,deprit1982delaunay,cid,abad2001short,abad2020integration,mahajan2018exact}). This method, for instance, was used to simplify the problem by an elimination of the parallax (\cite{deprit1981elimination}) which consisted on the removal of the short period terms from the system. Later, \cite{perigee} used a similar idea but to remove the long period terms associated with the evolution of the perigee, which provided several advantages in the computation of the solution (\cite{lara2019new}). 

In this work we introduce a formulation to describe the dynamical system of an object orbiting an oblate celestial body. Previous works that dealt with the definition of new set of elements focused on the removal of singularities and the simplification of the perturbation methodologies. This work departs from the past literature by seeking an element set that allows for a completely polynomial formulation, with the goal of allowing for new perturbation methods to be applied to the central gravity problem. To this end, our goal is to find a near-linear and completely polynomial system, which authors believe has not been tried before in the literature. This property provides several advantages when used with different perturbation techniques, including simplicity and lower computation load. This new formulation is achieved using an extended state-space of eight variables but it allows for a more direct application of perturbation methods while avoiding intermediary transformations to generate an analytical solution. Moreover, and in order to simplify the formulation, this work makes use of a universal variable based on a Sundman transformation (\cite{vallado2001fundamentals}). Additionally, these proposed state variables form a set of non-singular orbital elements that fully represents the zonal harmonics problem around an oblate celestial body, and that can be applied to any kind of orbit, including elliptic, parabolic, and hyperbolic orbits.

The application of this new formulation shows that the resultant system of differential equations represents the dynamic of two oscillators that have different frequencies but for the cases of unperturbed dynamics or periodic orbits. In that regard, we also show that these oscillators are decoupled in the unperturbed dynamic and coupled in the perturbed one. The presence of these two oscillators in the system motives the use of the Poncar\'e-Lindstedt perturbation method (\cite{birkhoff1927dynamical}) which is commonly used to solve systems of oscillators. Indeed, this work successfully applies a first order Poncar\'e-Lindstedt method to this formulation, obtaining the two natural frequencies of the dynamic, as well as the evolution of the orbital elements introduced in this manuscript, being this work the first direct application of the Poncar\'e-Lindstedt method to the central gravity problem known by the authors. This result is then compared with a numerical integration based on a Runge-Kutta scheme (\cite{davis2007methods}). Particularly, we show that the use of this analytical solution provides maximum errors in the position of the orbiting object of the order of tens of meters even for the cases of highly eccentric orbits around the Earth.

This manuscript is organized as follows. First, we introduce the new set of orbital elements and apply them to the non-perturbed problem. This will allow us to show a clear definition of these orbital elements as well as their relation with the more common keplerian or classical elements. Then, we present the solution of the linear system resultant of using this new set of orbital elements. Once the non-perturbed dynamic is shown, the perturbed problem with $J_2$ is studied, showing also how to transform the system of differential equations into one completely polynomial. After that, we follow the same approach but for the general case of the zonal harmonics of any order from the gravitational potential. Once the formulation is presented, we apply the Poncar\'e-Lindstedt method to this system to obtain a first order solution. Several examples of application of this solution are included to show the performance of the formulation and methodology. Finally, we present the non-dimensional formulation of the system resultant of using these orbital elements.


\section{Non-perturbed dynamic}

We start this manuscript with the presentation of the proposed set of variables applied to the non-perturbed problem. This is done in order to clearly show the definition and properties of this new set of variables and how they relate with the classical variables: the semi-major axis ($a$), the inclination ($i$), the eccentricity ($e$), the argument of perigee ($\omega$), the right ascension of the ascending node ($\Omega$), and the true anomaly ($\nu$).

In this work we introduce a new set of variables based on spherical coordinates. Therefore, we start this study by defining the problem in terms of spherical variables. Let $r$ be the distance from the orbiting object to the center of the celestial body, and $\varphi$ and $\lambda$ the latitude and longitude of the object with respect to the inertial frame of reference in a given instant of time. In these variables, the Hamiltonian of the system is:
\begin{equation}\label{eq:Hamiltonian_Kepler}
\mathcal{H} = \displaystyle\frac{1}{2}\left(p_r^2 + \frac{p_{\varphi}^2}{r^2} + \frac{p_{\lambda}^2}{r^2\cos^2(\varphi)}\right) - \frac{\mu}{r},
\end{equation}
where:
\begin{eqnarray}
p_r & = & \dot{r}; \nonumber \\
p_{\varphi} & = & r^2\dot{\varphi}; \nonumber \\
p_{\lambda} & = & r^2\cos^2(\varphi)\dot{\lambda};
\end{eqnarray}
are the conjugate momenta of the coordinates $r$, $\varphi$, and $\lambda$ respectively; $\mu$ is the gravitational constant of the celestial body; and where we denote $\dot{x}$ as the derivative with respect to time of a variable $x$. From this Hamiltonian, it is possible to obtain the Hamilton equations associated with the system:

\begin{eqnarray}\label{eq:spherical_nonperturbed}
\displaystyle\frac{dr}{dt} & = & p_r; \nonumber \\
\displaystyle\frac{dp_r}{dt} & = & -\frac{\mu}{r^2} + \frac{p_{\varphi}^2}{r^3}  + \frac{p_{\lambda}^2}{r^3\cos^2(\varphi)}; \nonumber \\
\displaystyle\frac{d\varphi}{dt} & = & \frac{p_{\varphi}}{r^2}; \nonumber \\
\displaystyle\frac{dp_{\varphi}}{dt} & = & -\frac{p_{\lambda}^2}{r^2}\frac{\sin(\varphi)}{\cos^3(\varphi)}; \nonumber \\
\displaystyle\frac{d\lambda}{dt} & = & \frac{p_{\lambda}}{r^2\cos^2(\varphi)}; \nonumber \\
\displaystyle\frac{dp_{\lambda}}{dt} & = & 0.
\end{eqnarray}
As it can be seen, this system of differential equations is highly non-linear, which limits its use in many applications, including the development of a perturbation theory based on this formulation. Therefore, the first objective is to obtain a linear system of equations by performing a series of variable transformations and a time regularization.


\subsection{Generation of a linear system}

\subsubsection{Definition of some first integrals}

In order to generate a linear system from the set of differential equations provided by Eq.~\eqref{eq:spherical_nonperturbed}, we require to find some of the first integrals of the system since we will make use of them later to derive all the new set of variables. In that regard, there are two first integrals that are directly obtained. The first one corresponds to the Hamiltonian $\mathcal{H}$ itself since the system is conservative. The second one is derived directly from the differential equations, where we see that $p_{\lambda}$ is a constant during the dynamic of the system, and thus, our second first integral. 

The third one can be obtained by relating the derivatives of $\varphi$ and $p_{\varphi}$:
\begin{equation}
\displaystyle\frac{d\varphi}{dp_{\varphi}} = \frac{p_{\varphi}}{p_{\lambda}^2}\frac{\cos^3(\varphi)}{\sin(\varphi)},
\end{equation}
which leads to:
\begin{equation}\label{eq:ptheta}
C_1 = \displaystyle\frac{p_{\varphi}^2}{p_{\lambda}^2} - \frac{1}{\cos^2(\varphi)},
\end{equation}
where $C_1$ is a constant of motion. However, instead of using $C_1$, we define the variable $p_{\theta}$ as $p_{\theta} = p_{\lambda}\displaystyle\sqrt{C_1}$, or in terms of the spherical elements:
\begin{equation}
p_{\theta}^2 = p_{\varphi}^2 - \displaystyle\frac{p_{\lambda}^2}{\cos^2(\varphi)}.
\end{equation}
It is important to note that $p_{\theta}$ is in fact the angular momentum of the orbit, which is also a constant quantity since $p_{\lambda}$ and $C_1$ are constants.

A fourth first integral can be obtained by relating the derivatives of $\varphi$ and $\lambda$:
\begin{equation}
\displaystyle\frac{d\varphi}{dp_{\lambda}} = \frac{p_{\varphi}}{p_{\lambda}}\cos^2(\varphi),
\end{equation}
which can be rewritten into a closed form integral by using Eq.~\eqref{eq:ptheta} and performing a change of variable from $\varphi$ to $\tan(\varphi)$:
\begin{equation}
\displaystyle\frac{d(\tan(\varphi))}{d\lambda} = \displaystyle\sqrt{\left(\displaystyle\frac{p_{\theta}}{p_{\lambda}}\right)^2 - 1 - \tan^2(\varphi)},
\end{equation}
whose integral is:
\begin{equation} \label{eq:beta}
\beta = \lambda - \arcsin\left(\displaystyle\frac{\tan(\varphi)}{\displaystyle\sqrt{\left(\displaystyle\frac{p_{\theta}}{p_{\lambda}}\right)^2 - 1}}\right),
\end{equation}
where $\beta$ is a constant of motion. Moreover, and in order to simplify the expression, we can define $\xi$ as:
\begin{equation} \label{eq:xi}
\xi = \displaystyle\frac{1}{\displaystyle\frac{p_{\theta}^2}{p_{\lambda}^2} - 1} = \frac{p_{\lambda}^2}{p_{\theta}^2 - p_{\lambda}^2},
\end{equation}
where as it can be seen, $\xi$ is also a constant of motion. That way, $\beta$ can be rewritten as:
\begin{equation}
\beta = \lambda - \arcsin\left(\tan(\varphi)\sqrt{\xi}\right).
\end{equation}
Nevertheless, it can be seen that for the same orbit, and depending on the initial conditions, we can have different values of $\beta$. For instance, in the nodes of an orbit the value of $\beta$ can be either $\Omega$ or $\Omega+\pi$. In order to solve this issue, we can define $\beta$ as: 
\begin{equation} \label{eq:beta_complete}
\beta = \begin{cases} \lambda - \arcsin\left(\tan(\varphi)\sqrt{\xi}\right) &\mbox{if } p_{\varphi} \geq 0 \\
\lambda + \arcsin\left(\tan(\varphi)\sqrt{\xi}\right) + \pi & \mbox{if } p_{\varphi} < 0 \end{cases},
\end{equation}
to avoid the duplicates generated by the $\arcsin$ function. That way, $\beta$ is defined as the right ascension of the ascending node $\Omega$ of the orbit (the longitude when the orbiting object is over the celestial body Equator and moving from South to North, also known as ascending pass).

Finally, we define an additional first integral that is dependent from the ones presented in Eqs.~\eqref{eq:ptheta} and~\eqref{eq:beta}, but that will be used later. This first integral can be obtained by relating the derivatives of $p_{\varphi}$ and $\lambda$:
\begin{equation}
\displaystyle\frac{dp_{\varphi}}{dp_{\lambda}} = -p_{\lambda}\tan(\varphi).
\end{equation}
Then, by using Eq.~\eqref{eq:ptheta} we can rewrite the equation in terms of $p_{\varphi}$ to be able to perform the integration:
\begin{equation}
\displaystyle\frac{1}{\cos^2(\varphi)} = 1 + \tan^2(\varphi) = \left(\frac{p_{\theta}}{p_{\lambda}}\right)^2 - \left(\frac{p_{\varphi}}{p_{\lambda}}\right)^2 
\end{equation}
therefore:
\begin{equation}
\tan(\varphi) = \sqrt{\left(\frac{p_{\theta}}{p_{\lambda}}\right)^2 - \left(\frac{p_{\varphi}}{p_{\lambda}}\right)^2 - 1} 
\end{equation}
and introducing this result in the previous equation leads to the following differential equation:
\begin{equation}
\displaystyle\frac{dp_{\varphi}}{dp_{\lambda}} = -p_{\lambda}\sqrt{\left(\frac{p_{\theta}}{p_{\lambda}}\right)^2 - \left(\frac{p_{\varphi}}{p_{\lambda}}\right)^2 - 1},
\end{equation}
whose solution is:
\begin{equation} \label{eq:eta}
\eta = \lambda - \arccos\left(\displaystyle\frac{p_{\varphi}}{p_{\lambda}}\frac{1}{\displaystyle\sqrt{\left(\displaystyle\frac{p_{\theta}}{p_{\lambda}}\right)^2 - 1}}\right) = \lambda - \arccos\left(\displaystyle\frac{p_{\varphi}}{p_{\lambda}}\sqrt{\xi}\right),
\end{equation}
where $\eta$ is a constant of motion. Note that as in the case of variable $\beta$, $\eta$ is not unique for each orbit due to the $\arccos$ function, and therefore, needs disambiguation.

\subsubsection{Definition of the new set of variables}

The objective now is to find a variable transformation that allows to generate a linear system. To that end, we first require to perform a time regularization based on an universal variable, also called Sundman transformation. We define, $\theta$ as a variable such that its time derivative is equal to:
\begin{equation} \label{eq:time_regularization}
\displaystyle\frac{d\theta}{dt} = \frac{p_{\theta}}{r^2}.
\end{equation}
As it can be observed, this new variable $\theta$ is related with the angular momentum of the orbit $p_{\theta}$, and can be identified with the argument of latitude (or, alternatively, the true anomaly) of the orbit if the constants of integration are properly chosen for the unperturbed problem. Therefore, using $\theta$ as the new independent variable, the system of differential equations from Eq.~\eqref{eq:spherical_nonperturbed} becomes:

\begin{eqnarray}\label{eq:spherical_nonperturbed_regularized}
\displaystyle\frac{dr}{d\theta} & = & \frac{p_r}{p_{\theta}}r^2; \nonumber \\
\displaystyle\frac{dp_r}{d\theta} & = & -\frac{\mu}{p_{\theta}} + \frac{p_{\varphi}^2}{rp_{\theta}}  + \frac{p_{\lambda}^2}{r\cos^2(\varphi)p_{\theta}}; \nonumber \\
\displaystyle\frac{d\varphi}{d\theta} & = & \frac{p_{\varphi}}{p_{\theta}}; \nonumber \\
\displaystyle\frac{dp_{\varphi}}{d\theta} & = & -\frac{p_{\lambda}^2}{p_{\theta}}\frac{\sin(\varphi)}{\cos^3(\varphi)}; \nonumber \\
\displaystyle\frac{d\lambda}{d\theta} & = & \frac{p_{\lambda}}{p_{\theta}\cos^2(\varphi)}; \nonumber \\
\displaystyle\frac{dp_{\lambda}}{d\theta} & = & 0.
\end{eqnarray}

Once the time regularization is performed, we perform a variable transformation to make the system completely linear. In particular, we propose $\alpha$, $p_r$, $s$, $\gamma$, $\beta$, $p_{\lambda}$ as the new set of variables in the problem, where $\alpha$, $s$, and $\gamma$ are defined as:
\begin{eqnarray} \label{eq:basic_transform}
\alpha & = & \displaystyle\frac{p_{\theta}}{r} - \frac{\mu}{p_{\theta}} = \frac{p_{\varphi}^2}{rp_{\theta}}  + \frac{p_{\lambda}^2}{r\cos^2(\varphi)p_{\theta}} - \frac{\mu}{p_{\theta}}, \nonumber \\ 
s & = & \sin(\varphi), \nonumber \\
\gamma & = & \displaystyle\frac{p_{\varphi}}{p_{\theta}}\cos(\varphi).
\end{eqnarray}
and whose inverse transformations are:
\begin{eqnarray}
r & = & \displaystyle\frac{p_{\lambda}^2}{p_{\lambda}\alpha\sqrt{1-s^2-\gamma^2}+\mu\left(1-s^2-\gamma^2\right)}, \nonumber \\ 
\varphi & = & \arcsin(s), \nonumber \\
p_{\varphi} & = & p_{\lambda}\displaystyle\frac{\gamma}{\cos(\varphi)}\sqrt{\frac{1}{1-s^2-\gamma^2}}.
\end{eqnarray}
Therefore, we have to derive which are the derivatives of this new set of variables with respect to $\theta$. The derivative of $\alpha$ is provided by:
\begin{eqnarray}
\displaystyle\frac{d\alpha}{d\theta} = -\displaystyle\frac{p_{\theta}}{r^2}\frac{dr}{d\theta} = -\frac{p_{\theta}}{r^2}\frac{p_r}{p_{\theta}}r^2 = -p_r. 
\end{eqnarray}
On the other hand, the derivative of $s$ is:
\begin{equation}
\displaystyle\frac{d s}{d\theta} = \cos(\varphi)\frac{d\varphi}{d\theta} = \frac{p_{\varphi}}{p_{\theta}}\cos(\varphi) = \gamma.
\end{equation}
Finally, the derivative of $\gamma$ can be obtained through the following equation:
\begin{eqnarray}
\displaystyle\frac{d\gamma}{d\theta} & = & \displaystyle\frac{1}{p_{\theta}}\cos(\varphi)\frac{dp_{\varphi}}{d\theta} - \frac{p_{\varphi}}{p_{\theta}}\sin(\varphi)\frac{d\varphi}{d\theta} \nonumber \\
& = & -\frac{1}{p_{\theta}}\cos(\varphi)\frac{p_{\lambda}^2}{p_{\theta}}\frac{\sin(\varphi)}{\cos^3(\varphi)} - \frac{p_{\varphi}}{p_{\theta}}\sin(\varphi)\frac{p_{\varphi}}{p_{\theta}},
\end{eqnarray}
where we can use Eq.~\eqref{eq:ptheta} to simplify the expression:
\begin{equation}
\displaystyle\frac{d\gamma}{d\theta} = -\sin(\varphi)\left[\frac{p_{\lambda}^2}{p_{\theta}^2}\frac{1}{\cos^2(\varphi)} - \frac{p_{\varphi}^2}{p_{\theta}^2}\right] = -\sin(\varphi) = -s.
\end{equation}
Therefore, using $\alpha$, $p_r$, $s$, $\gamma$, $\beta$, and $p_{\lambda}$ as the new set of orbital variables allows to transform the nonlinear system of differential equations from Eq.\eqref{eq:spherical_nonperturbed_regularized} into this linear system:
\begin{eqnarray}\label{eq:spherical_nonperturbed_regularized_final}
\displaystyle\frac{d\alpha}{d\theta} & = & -p_r; \nonumber \\
\displaystyle\frac{dp_r}{d\theta} & = & \alpha; \nonumber \\
\displaystyle\frac{d s}{d\theta} & = & \gamma; \nonumber \\
\displaystyle\frac{d\gamma}{d\theta} & = & -s; \nonumber \\
\displaystyle\frac{d\beta}{d\theta} & = & 0; \nonumber \\
\displaystyle\frac{dp_{\lambda}}{d\theta} & = & 0.
\end{eqnarray}

Using these variables and first integrals, the Hamiltonian, and thus, the energy of the system can be expressed as:
\begin{equation} \label{eq:new_hamiltonian}
\mathcal{H} = \displaystyle\frac{1}{2}\left(p_r^2 + \alpha^2 - \frac{\mu^2}{p_{\theta}^2}\right).
\end{equation} 
Also, it is worth noticing that for equatorial orbits, $\beta$ is not defined. However, in these cases the derivative of $\lambda$ from Eq.~\eqref{eq:spherical_nonperturbed_regularized} becomes 1, and thus, the evolution of the longitude of the orbit is $\lambda = \theta$.

\subsection{Solution of the linear system}

Equation~\eqref{eq:spherical_nonperturbed_regularized_final} provides a complete linear system of differential equations that can be solved for any initial condition. Let $\alpha_0$, $p_{r0}$, $s_{0}$, $\gamma_0$, $\beta=\beta_0$, and $p_{\lambda}=p_{\lambda 0}$ be the initial conditions of the problem. Then, from the first two differential equations from Eq.~\eqref{eq:spherical_nonperturbed_regularized_final} we can obtain the following general solution for $\alpha$ and $p_r$:
\begin{eqnarray}
\alpha & = & \alpha_0\cos(\theta-\theta_0) - p_{r0}\sin(\theta-\theta_0), \nonumber \\
p_r & = & \alpha_0\sin(\theta-\theta_0) + p_{r0}\cos(\theta-\theta_0),
\end{eqnarray}
where $\theta_0$ and $\theta$ are the initial and final arguments of latitude of the motion.
From the solution in $\alpha$ it is possible to obtain the evolution of the radius of the orbit:
\begin{eqnarray}
r & = & \displaystyle\frac{p_{\theta}}{\alpha + \displaystyle\frac{\mu}{p_{\theta}}} = \displaystyle\frac{p_{\theta}}{\alpha_0\cos(\theta-\theta_0) - p_{r0}\sin(\theta-\theta_0) + \displaystyle\frac{\mu}{p_{\theta}}} \nonumber \\
& = & \displaystyle\frac{p_{\theta}^2/\mu}{1 + \frac{\alpha_0p_{\theta}}{\mu}\cos(\theta-\theta_0) - \frac{p_{r0}p_{\theta}}{\mu}\sin(\theta-\theta_0)},
\end{eqnarray}
which represents a more general expression to the well known relation between the radial distance ($r$) and the true anomaly ($\nu$) of an elliptic orbit:
\begin{equation} \label{eq:kepler_r}
r = \displaystyle\frac{a\left(1 - e^2\right)}{1 + e\cos(\nu)},
\end{equation}
where $a$ and $e$ are the semi-major axis and eccentricity of the orbit.

In the same way, the solution for $s$ and $\gamma$ can be obtained:
\begin{eqnarray} \label{eq:s_gamma}
s & = & s_0\cos(\theta-\theta_0) + \gamma_0\sin(\theta-\theta_0), \nonumber \\
\gamma & = & - s_0\sin(\theta-\theta_0) + \gamma_0\cos(\theta-\theta_0),
\end{eqnarray}
and from them, the spherical variables $\varphi$ and $p_{\varphi}$:
\begin{eqnarray}
\varphi & = & \arcsin(s), \nonumber \\
p_{\varphi} & = & \displaystyle\frac{\gamma}{\cos(\varphi)}p_{\theta}.
\end{eqnarray}
In this regard, it is important to note that for polar orbits, $p_{\varphi} = p_{\theta}$, and thus, $\gamma = \cos(\varphi)$ $\forall \varphi$, so there is no singularity in the poles. In addition, since the range in the latitude of the orbit is $\varphi\in[-\pi/2,\pi/2]$ the $\arcsin(\varphi)$ is a bijective function in the entire domain. Therefore, these transformations are well defined no matter the type of orbit or the position of the orbiting object. 

The longitude of the orbit, on the other hand, is obtained using the previous solutions of $s$ and $\gamma$ from Eq.~\eqref{eq:s_gamma} and Eq.~\eqref{eq:beta}. However, there is one important consideration to be made. The definition of $\beta$ in Eq.~\eqref{eq:beta} contains an arcsine as a function of $\varphi$. This means that the transformation from $\{\varphi,p_{\varphi}\}$ to $\lambda$ is not completely well defined since it can result in two different values. This issue can be solved for instance by using the values of $\beta$ and $\eta$:
\begin{eqnarray}
\sin(\lambda) & = & \sin\left(\beta + \arcsin\left(\tan(\varphi)\sqrt{\xi}\right)\right), \nonumber \\
\cos(\lambda) & = & \cos\left(\eta + \arccos\left(\displaystyle\frac{p_{\varphi}}{p_{\lambda}}\sqrt{\xi}\right)\right),
\end{eqnarray}
where the longitude can be disambiguated using the arctan2 function. This expression can be only applied in prograde orbits. However for retrograde orbits, the square root inside the $\arcsin$ function changes sign, and then, in those cases, the expressions become:
\begin{eqnarray}
\sin(\lambda) & = & \sin\left(\beta - \arcsin\left(\tan(\varphi)\sqrt{\xi}\right)\right), \nonumber \\
\cos(\lambda) & = & \cos\left(\eta + \arccos\left(\displaystyle\frac{p_{\varphi}}{p_{\lambda}}\sqrt{\xi}\right)\right),
\end{eqnarray}
Alternatively, we can perform an equivalent derivation using just $\beta$ and the sign of $p_{\varphi}$ to determine the orientation of the movement, that is:
\begin{equation}
\lambda = \begin{cases} \beta + \arcsin\left(\tan(\varphi)\sqrt{\xi}\right) &\mbox{if } p_{\varphi} \geq 0 \\
\beta - \arcsin\left(\tan(\varphi)\sqrt{\xi}\right) + \pi & \mbox{if } p_{\varphi} < 0 \end{cases},
\end{equation}
for prograde orbits and:
\begin{equation}
\lambda = \begin{cases} \beta - \arcsin\left(\tan(\varphi)\sqrt{\xi}\right) &\mbox{if } p_{\varphi} \geq 0 \\
\beta + \arcsin\left(\tan(\varphi)\sqrt{\xi}\right) + \pi & \mbox{if } p_{\varphi} < 0 \end{cases},
\end{equation}
for retrograde orbits.

Therefore, the only thing left to solve is the time evolution of the system. From the time regularization from Eq.~\eqref{eq:time_regularization} we can define the derivative of time with respect to $\theta$:
\begin{equation}
\displaystyle\frac{d\theta}{dt} = \frac{p_{\theta}}{r^2} = \displaystyle\frac{\left(1 + \displaystyle\frac{\alpha_0p_{\theta}}{\mu}\cos(\theta-\theta_0) - \frac{p_{r0}p_{\theta}}{\mu}\sin(\theta-\theta_0)\right)^2}{p_{\theta}^3/\mu^2}.
\end{equation} 
In addition, since $r$ and $p_{\theta}$ can be written as a function of $\theta$, the direct integration of the previous equation can be performed to obtain the following relation:
\begin{eqnarray} \label{eq:time_sol}
t & = & t_0 + \displaystyle\frac{\frac{p_{\theta}}{2\mathcal{H}}\left(\alpha_0^2+\alpha_0\frac{\mu}{p_{\theta}}+p_{r0}^2\right) \sin\left(\theta-\theta_0\right)}{(\alpha_0+\frac{\mu}{p_{\theta}})  (\frac{\mu}{p_{\theta}}+\alpha_0 \cos\left(\theta-\theta_0\right)-p_{r0}
	\sin\left(\theta-\theta_0\right))} \nonumber \\
& - & \displaystyle\frac{p_{\theta}}{2\mathcal{H}}\displaystyle\frac{\frac{\mu p_{r0}}{p_{\theta}} \left(1-\cos\left(\theta-\theta_0\right)\right)}{(\alpha_0+\frac{\mu}{p_{\theta}})  (\frac{\mu}{p_{\theta}}+\alpha_0 \cos\left(\theta-\theta_0\right)-p_{r0}
	\sin\left(\theta-\theta_0\right))} \nonumber \\
& + & \displaystyle\frac{2\mu}{\left(2\mathcal{H}\right)^{3/2}} \text{artanh}\left(\frac{p_{r0}}{\sqrt{2\mathcal{H}}}\right)  \nonumber \\
& - & \displaystyle\frac{2\mu}{\left(2\mathcal{H}\right)^{3/2}}\text{artanh}\left(\frac{p_{r0}+(\alpha_0-\frac{\mu}{p_{\theta}})\tan\left(\frac{\theta - \theta_0}{2}\right)}{\sqrt{2\mathcal{H}}}\right),
\end{eqnarray}
which relates the time evolution with the argument of latitude of the orbit. Nevertheless, there is an important thing to take into account when using this expression. Since the expression provided by Eq.~\eqref{eq:time_sol} only depends on the angle $\theta$, the result of the time evolution ranges in $(t-t_0) \in[-T/2,T/2]$, where $T$ is the period of the orbit. This means that in order to know the time of propagation it is also required to take into account the number of complete orbital revolutions performed.

\subsection{Relation with classical variables}

The objective of this section is to relate the set of variables proposed in this work with the classical elements semi-major axis ($a$), eccentricity ($e$), inclination ($i$), argument of perigee ($\omega$), right ascension of the ascending node ($\Omega$), and true anomaly ($\nu$). To that end, we show in the following subsections both the direct and inverse transformations associated with the set of variables proposed in this work. 

\subsubsection{From spherical to classical elements}

We know that the Hamiltonian of an orbit in classical elements is:
\begin{equation}
\mathcal{H} = -\displaystyle\frac{\mu}{2a},
\end{equation}
which is equivalent to the Hamiltonian from Eq.~\eqref{eq:new_hamiltonian}, that is:
\begin{equation}
\mathcal{H} = -\displaystyle\frac{\mu}{2a} = \frac{1}{2}\left(p_r^2 + \alpha^2 - \frac{\mu^2}{p_{\theta}^2}\right).
\end{equation}
Therefore, the semi-major axis of the orbit is:
\begin{equation} \label{eq:semi-major_axis}
a = \displaystyle\frac{\mu}{\frac{\mu^2}{p_{\theta}^2} - p_r^2 - \alpha^2}.
\end{equation}
On the other hand, we know that the angular momentum of the orbit can be expressed in classical elements as:
\begin{equation} \label{eq:kepler_ang_momentum}
p_{\theta} = \displaystyle\sqrt{\mu a\left(1-e^2\right)},
\end{equation}
and using the result form Eq.~\eqref{eq:semi-major_axis} we can obtain the value of the eccentricity:
\begin{equation} \label{eq:kepler_eccen}
e = \displaystyle\frac{p_{\theta}}{\mu}\displaystyle\sqrt{p_r^2 + \alpha^2}.
\end{equation}

The inclination can be obtained using the projection of the angular momentum in the $z$-axis:
\begin{equation} \label{eq:kepler_longitude_momentum}
p_{\lambda} = p_{\theta}\cos(i),
\end{equation}
and so:
\begin{equation}
i = \arccos\left(\displaystyle\frac{p_{\lambda}}{p_{\theta}}\right).
\end{equation}
Alternatively, the inclination can be obtained through:
\begin{eqnarray} \label{eq:sin2inc}
\sin^2(i) & = & 1 - \cos^2(i) = 1 - \displaystyle\frac{p_{\lambda}^2}{p_{\theta}^2} = 1 - \cos^2(\varphi)\left(1 - \frac{p_{\varphi}^2}{p_{\theta}^2}\right) \nonumber \\
& = & \sin^2(\varphi) + \left(\frac{p_{\varphi}}{p_{\theta}}\cos(\varphi)\right)^2 = s^2 + \gamma^2,
\end{eqnarray}
and thus:
\begin{equation}
i = \arcsin\left(\sqrt{s^2+\gamma^2}\right).
\end{equation}

In order to obtain the value of the right ascension of the ascending node, we first need to determine the variables $\lambda$, $\varphi$, $p_{\varphi}$ and $\xi$ from the expressions presented in the previous subsection. Once this variables are available, the value of the right ascension of the ascending node can be obtained through the first integral $\beta$. In particular, we have to impose the condition that the longitude of the orbit at the ascending pass over the Equator is in fact the right ascension of the ascending node. That way:
\begin{equation}
\Omega = \beta = \begin{cases} \lambda - \arcsin\left(\tan(\varphi)\sqrt{\xi}\right) &\mbox{if } p_{\varphi} \geq 0 \\
\lambda + \arcsin\left(\tan(\varphi)\sqrt{\xi}\right) + \pi & \mbox{if } p_{\varphi} < 0 \end{cases}.
\end{equation}

Finally, the true anomaly can be obtained using Eq.~\eqref{eq:kepler_r}:
\begin{equation}
	\nu =  \begin{cases} \arccos\left(\displaystyle\frac{a\left(1-e^2\right)}{er} - \frac{1}{e}\right) &\mbox{if } p_r \geq 0 \\
	2\pi - \arccos\left(\displaystyle\frac{a\left(1-e^2\right)}{er} - \frac{1}{e}\right) & \mbox{if } p_r < 0 \end{cases}
\end{equation}
On the other hand, the argument of latitude $u = \omega + \nu$ of the orbit in a given instant can be derived using spherical trigonometry:
\begin{equation}
	u = \begin{cases} \arccos\left(\cos(\lambda-\Omega)\cos(\varphi)\right) &\mbox{if } \varphi \geq 0 \\
	2\pi -\arccos\left(\cos(\lambda-\Omega)\cos(\varphi)\right) & \mbox{if } \varphi < 0 \end{cases},
\end{equation}
and with it, the argument of perigee of the orbit can be obtained $\omega = u - \nu$.

\subsubsection{From classical to spherical elements}

The radial distance can be easily obtained through the well known Eq.~\eqref{eq:kepler_r} while the angular momentum with Eq.~\eqref{eq:kepler_ang_momentum}. With $r$ and $p_{\theta}$ we can obtain the value of $\alpha$ using the definition from Eq.~\eqref{eq:basic_transform}. Then, from Eq.~\eqref{eq:kepler_eccen} and using the geometry of the orbit:
\begin{equation}
p_r = \begin{cases} + \sqrt{\frac{\mu^2}{p_{\theta}^2}e^2 - \alpha^2} &\mbox{if } \sin(\nu) \geq 0 \\
- \sqrt{\frac{\mu^2}{p_{\theta}^2}e^2 - \alpha^2} & \mbox{if } \sin(\nu) < 0 \end{cases}.
\end{equation}
Also, the conjugate momenta of the longitude $p_{\lambda}$ can be obtained directly using Eq.~\eqref{eq:kepler_longitude_momentum}, and this result used to obtain $\xi$ using Eq.~\eqref{eq:xi}. On the other hand, we know that $\beta = \Omega$ when considering the definition of $\beta$ provided by Eq.~\eqref{eq:beta_complete} to avoid duplicates in the formulation. 

From the argument of latitude of the orbit $u = \omega+\nu$ we can obtain the orbital element $s$ using spherical trigonometry:
\begin{equation}
s = \sin(\varphi) = \sin(i)\sin(\omega+\nu).
\end{equation}
Finally, from Eq.~\eqref{eq:sin2inc}, the value of $\gamma$ can be obtained:
\begin{equation}
\gamma = \begin{cases} + \sqrt{\sin^2(i) - s^2} &\mbox{if } \cos(\omega + \nu) \geq 0 \\
- \sqrt{\sin^2(i) - s^2} & \mbox{if } \cos(\omega + \nu) < 0 \end{cases}.
\end{equation}


\section{J2 formulation}

Once the non-perturbed formulation of the set of variables proposed in this study is presented, we start with the assessment of orbital perturbations. In particular, in this section we focus on the effect of the oblateness of a celestial body, represented by the $J_2$ term of the gravitational potential. Therefore, we know that the Hamiltonian in spherical coordinates of this perturbed problem with $J_2$ is:
\begin{equation}
\mathcal{H} = \displaystyle\frac{1}{2}\left(p_r^2 + \frac{p_{\varphi}^2}{r^2} + \frac{p_{\lambda}^2}{r^2\cos^2(\varphi)}\right) - \frac{\mu}{r} + \frac{1}{2}\mu R^2_{\oplus}J_2\frac{1}{r^3}\left(3\sin^2(\varphi) - 1\right),
\end{equation}
where $R_{\oplus}$ is the radius of the celestial body at the Equator. From this Hamiltonian, the associated Hamilton equations can be obtained:
\begin{eqnarray}
\displaystyle\frac{dr}{dt} & = & p_r; \nonumber \\
\displaystyle\frac{dp_r}{dt} & = & -\frac{\mu}{r^2} + \frac{p_{\varphi}^2}{r^3}  + \frac{p_{\lambda}^2}{r^3\cos^2(\varphi)} + \frac{3}{2}\mu J_2 R_{\oplus}^2\frac{1}{r^4}\left(3\sin^2(\varphi) - 1\right); \nonumber \\
\displaystyle\frac{d\varphi}{dt} & = & \frac{p_{\varphi}}{r^2}; \nonumber \\
\displaystyle\frac{dp_{\varphi}}{dt} & = & -\frac{p_{\lambda}^2}{r^2}\frac{\sin(\varphi)}{\cos^3(\varphi)} - 3\mu J_2 R_{\oplus}^2\frac{1}{r^3}\sin(\varphi)\cos(\varphi); \nonumber \\
\displaystyle\frac{d\lambda}{dt} & = & \frac{p_{\lambda}}{r^2\cos^2(\varphi)}; \nonumber \\
\displaystyle\frac{dp_{\lambda}}{dt} & = & 0.
\end{eqnarray}
In this system of differential equations we can perform the same set of transformations as in the unperturbed problem, including the time regularization from Eq.~\eqref{eq:time_regularization}, and changes of variables provided by Eqs.~\eqref{eq:beta} and~\eqref{eq:basic_transform} to obtain:
\begin{eqnarray}\label{eq:j2_diferential}
\displaystyle\frac{d\alpha}{d\theta} & = & -p_r - 3\mu J_2 R_{\oplus}^2\frac{\left(1-s^2-\gamma^2\right)^{3/2}}{p_{\lambda}^3}\nonumber \\
& & \left(\alpha+\mu\frac{\displaystyle\sqrt{1-s^2-\gamma^2}}{p_{\lambda}}\right)\left(\alpha+2\mu\frac{\displaystyle\sqrt{1-s^2-\gamma^2}}{p_{\lambda}}\right)\gamma s; \nonumber \\
\displaystyle\frac{dp_r}{d\theta} & = & \alpha + \frac{3}{2}\mu J_2 R_{\oplus}^2\frac{\left(1-s^2-\gamma^2\right)^{3/2}}{p_{\lambda}^3} \left(\alpha+\mu\frac{\displaystyle\sqrt{1-s^2-\gamma^2}}{p_{\lambda}}\right)^2\left(3s^2-1\right); \nonumber \\
\displaystyle\frac{d s}{d\theta} & = & \gamma; \nonumber \\
\displaystyle\frac{d\gamma}{d\theta} & = & -s - 3\mu J_2 R_{\oplus}^2\frac{\left(1-s^2-\gamma^2\right)^{5/2}}{p_{\lambda}^3}\left(\alpha+\mu\frac{\displaystyle\sqrt{1-s^2-\gamma^2}}{p_{\lambda}}\right)s; \nonumber \\
\displaystyle\frac{d\beta}{d\theta} & = & - 3\mu J_2 R_{\oplus}^2\frac{\left(1-s^2-\gamma^2\right)^2}{p_{\lambda}^3\left(s^2+\gamma^2\right)}\left(\alpha+\mu\frac{\displaystyle\sqrt{1-s^2-\gamma^2}}{p_{\lambda}}\right)s^2; \nonumber \\
\displaystyle\frac{dp_{\lambda}}{d\theta} & = & 0,
\end{eqnarray}
which completely defines the dynamic of a orbiting object subjected to the $J_2$ perturbation.

The objective now is to generate an equivalent system of differential equations that is completely polynomial. In order to do that, we have to expand the space of configuration of the system by the inclusion of two new orbital elements $I_{\theta}$ and $\xi$. 

Let $I_{\theta}$ be a new orbital element defined as:
\begin{equation}
I_{\theta} = \displaystyle\frac{1}{p_{\theta}} = \frac{1}{p_{\lambda}}\displaystyle\sqrt{1-s^2-\gamma^2},
\end{equation}
and whose derivative with respect to $\theta$ is:
\begin{equation}
\displaystyle\frac{dI_{\theta}}{d\theta} = 3\mu J_2R_{\oplus}^2I_{\theta}^4\left(\alpha + \mu I_{\theta}\right)s\gamma.
\end{equation}
We introduce this orbital element into the system of differential equations from Eq.~\eqref{eq:j2_diferential}, which leads to:

\begin{eqnarray}
\displaystyle\frac{d\alpha}{d\theta} & = & -p_r - 3\mu J_2 R_{\oplus}^2I_{\theta}^3\left(\alpha+\mu I_{\theta}\right)\left(\alpha+2\mu I_{\theta}\right)\gamma s; \nonumber \\
\displaystyle\frac{dp_r}{d\theta} & = & \alpha + \frac{3}{2}\mu J_2 R_{\oplus}^2I_{\theta}^3\left(\alpha+\mu I_{\theta}\right)^2\left(3s^2-1\right); \nonumber \\
\displaystyle\frac{d s}{d\theta} & = & \gamma; \nonumber \\
\displaystyle\frac{d\gamma}{d\theta} & = & -s - 3\mu J_2 R_{\oplus}^2\left(1-s^2-\gamma^2\right)I_{\theta}^3\left(\alpha+\mu I_{\theta}\right)s \nonumber \\
& = & -s - 3\mu J_2 R_{\oplus}^2I_{\theta}^5p_{\lambda}^2\left(\alpha+\mu I_{\theta}\right)s; \nonumber \\
\displaystyle\frac{dI_{\theta}}{d\theta} & = & 3\mu J_2R_{\oplus}^2I_{\theta}^4\left(\alpha + \mu I_{\theta}\right)s\gamma; \nonumber \\
\displaystyle\frac{d\beta}{d\theta} & = & - 3\mu J_2 R_{\oplus}^2\frac{\left(1-s^2-\gamma^2\right)^2}{p_{\lambda}^3\left(s^2+\gamma^2\right)}\left(\alpha+\mu I_{\theta}\right)s^2; \nonumber \\
\displaystyle\frac{dp_{\lambda}}{d\theta} & = & 0,
\end{eqnarray}
As it can be seen, all the equations became polynomial except the derivative of $\beta$. In order to solve that, we introduce another orbital element $\xi$ defined as in Eq.~\eqref{eq:xi}:
\begin{equation}
\xi = \displaystyle\frac{p_{\lambda}^2}{p_{\theta}^2 - p_{\lambda}^2} = \frac{1-s^2-\gamma^2}{s^2+\gamma^2},
\end{equation}
and whose derivative is:
\begin{equation}
\displaystyle\frac{d\xi}{d\theta} = 6\mu J_2R_{\oplus}^2\xi^2\frac{I_{\theta}}{p_{\lambda}^2}\left(\alpha + \mu I_{\theta}\right)s\gamma.
\end{equation}
Introducing this new variable transforms the system of differential equations into:
\begin{eqnarray}\label{eq:j2_poly}
\displaystyle\frac{d\alpha}{d\theta} & = & -p_r - 3\mu J_2 R_{\oplus}^2I_{\theta}^3\left(\alpha+\mu I_{\theta}\right)\left(\alpha+2\mu I_{\theta}\right)\gamma s; \nonumber \\
\displaystyle\frac{dp_r}{d\theta} & = & \alpha + \frac{3}{2}\mu J_2 R_{\oplus}^2I_{\theta}^3\left(\alpha+\mu I_{\theta}\right)^2\left(3s^2-1\right); \nonumber \\
\displaystyle\frac{d s}{d\theta} & = & \gamma; \nonumber \\
\displaystyle\frac{d\gamma}{d\theta} & = & -s - 3\mu J_2 R_{\oplus}^2I_{\theta}^5p_{\lambda}^2\left(\alpha+\mu I_{\theta}\right)s; \nonumber \\
\displaystyle\frac{dI_{\theta}}{d\theta} & = & 3\mu J_2R_{\oplus}^2I_{\theta}^4\left(\alpha + \mu I_{\theta}\right)s\gamma; \nonumber \\
\displaystyle\frac{d\beta}{d\theta} & = & - 3\mu J_2 R_{\oplus}^2\xi\frac{1}{p_{\lambda}}I_{\theta}^2\left(\alpha+\mu I_{\theta}\right)s^2; \nonumber \\
\displaystyle\frac{d\xi}{d\theta} & = & 6\mu J_2R_{\oplus}^2\xi^2\frac{I_{\theta}}{p_{\lambda}^2}\left(\alpha + \mu I_{\theta}\right)s\gamma; \nonumber \\
\displaystyle\frac{dp_{\lambda}}{d\theta} & = & 0,
\end{eqnarray}
which as it can be seen, the system is now completely polynomial.

In addition, we can also introduce the variable $\eta$ (as defined in Eq.~\eqref{eq:eta}) as a new orbital element for the system to disambiguate the value of the longitude of the orbit. In that respect, the derivative of $\eta$ is:
\begin{equation}
\displaystyle\frac{d\eta}{d\theta} = 3\mu J_2R_{\oplus}^2\frac{1}{p_{\lambda}}\left(\alpha + \mu I_{\theta}\right)\left(\frac{\xi}{p_{\lambda}^2} - I_{\theta}^2(1-s^2)\right),
\end{equation}
which is also in polynomial form.


\section{Zonal formulation}

It is possible to extend the previous result to the zonal harmonics of any order. Particularly, the Hamiltonian for a perturbed dynamic up to zonal harmonics of order $m$ is:
\begin{equation}
\mathcal{H} = \displaystyle\frac{1}{2}\left(p_r^2 + \frac{p_{\varphi}^2}{r^2} + \frac{p_{\lambda}^2}{r^2\cos^2(\varphi)}\right) - \frac{\mu}{r}\left(1 - \sum_{n=2}^{m}J_nP_n\left(\sin(\varphi)\right)\frac{R_{\oplus}^n}{r^n}\right),
\end{equation}
where $P_n(\sin(\varphi))$ are the Legendre polynomials of order $n$ in the variable $\sin(\varphi)$. The associated Hamilton equations are then:
\begin{eqnarray}
\displaystyle\frac{dr}{dt} & = & p_r; \nonumber \\
\displaystyle\frac{dp_r}{dt} & = & -\frac{\mu}{r^2} + \frac{p_{\varphi}^2}{r^3}  + \frac{p_{\lambda}^2}{r^3\cos^2(\varphi)} \nonumber \\
& + & \sum_{n=2}^m (n+1)\mu J_n P_n(\sin(\varphi))\frac{R_{\oplus}^n}{r^{n+2}}; \nonumber \\
\displaystyle\frac{d\varphi}{dt} & = & \frac{p_{\varphi}}{r^2}; \nonumber \\
\displaystyle\frac{dp_{\varphi}}{dt} & = & -\frac{p_{\lambda}^2}{r^2}\frac{\sin(\varphi)}{\cos^3(\varphi)} - \sum_{n=2}^m \mu J_n \frac{\partial P_n(\sin(\varphi))}{\partial\varphi}\frac{R_{\oplus}^n}{r^{n+1}}; \nonumber \\
\displaystyle\frac{d\lambda}{dt} & = & \frac{p_{\lambda}}{r^2\cos^2(\varphi)}; \nonumber \\
\displaystyle\frac{dp_{\lambda}}{dt} & = & 0.
\end{eqnarray}
which performing the time regularization from Eq.~\eqref{eq:time_regularization} and the set of element transformations from Eqs.~\eqref{eq:basic_transform} and~\eqref{eq:beta}, leads to:

\begin{eqnarray}
\displaystyle\frac{d\alpha}{d\theta} & = & -p_r - \sum_{n=2}^{m}\mu J_nR_{\oplus}^n\displaystyle\frac{\partial P_n(s)}{\partial s}\gamma \frac{\left(1-s^2-\gamma^2\right)^{(n+1)/2}}{p_{\lambda}^{n+1}} \nonumber \\
& & \left(\alpha+\mu \frac{\sqrt{1-s^2-\gamma^2}}{p_{\lambda}}\right)^{n-1}\left(\alpha + 2\mu \frac{\sqrt{1-s^2-\gamma^2}}{p_{\lambda}}\right); \nonumber \\
\displaystyle\frac{dp_r}{d\theta} & = & \alpha + \sum_{n=2}^m \left(n+1\right)\mu J_n R_{\oplus}^nP_n(s)\frac{\left(1-s^2-\gamma^2\right)^{(n+1)/2}}{p_{\lambda}^{n+1}} \nonumber \\
& & \left(\alpha + \mu \frac{\sqrt{1-s^2-\gamma^2}}{p_{\lambda}}\right)^n; \nonumber \\
\displaystyle\frac{d s}{d\theta} & = & \gamma; \nonumber \\
\displaystyle\frac{d\gamma}{d\theta} & = & -s - \sum_{n=2}^m \mu J_n R_{\oplus}^n \frac{\partial P_n (s)}{\partial s} \frac{\left(1-s^2-\gamma^2\right)^{(n+3)/2}}{p_{\lambda}^{n+1}} \nonumber \\
& & \left(\alpha + \mu \frac{\sqrt{1-s^2-\gamma^2}}{p_{\lambda}}\right)^{n-1}; \nonumber \\
\displaystyle\frac{d \beta}{d\theta} & = & -\sum_{n=2}^m \mu J_n R_{\oplus}^n \frac{\partial P_n (s)}{\partial s} \frac{s}{s^2+\gamma^2} \frac{\left(1-s^2-\gamma^2\right)^{(n+2)/2}}{p_{\lambda}^{n+1}} \nonumber \\
& & \left(\alpha + \mu \frac{\sqrt{1-s^2-\gamma^2}}{p_{\lambda}}\right)^{n-1}; \nonumber \\
\displaystyle\frac{dp_{\lambda}}{d\theta} & = & 0.
\end{eqnarray}

Moreover, we can perform the same expansion using the variables $I_{\theta}$ and $\xi$ to transform the former expression into a polynomial system of differential equations:

\begin{eqnarray} \label{eq:complete_poly}
\displaystyle\frac{d\alpha}{d\theta} & = & -p_r - \sum_{n=2}^{m}\mu J_nR_{\oplus}^n\displaystyle\frac{\partial P_n(s)}{\partial s}\gamma I_{\theta}^{n+1}  \left(\alpha+\mu I_{\theta}\right)^{n-1} \left(\alpha + 2\mu I_{\theta}\right); \nonumber \\
\displaystyle\frac{dp_r}{d\theta} & = & \alpha + \sum_{n=2}^m \left(n+1\right)\mu J_n R_{\oplus}^nP_n(s)I_{\theta}^{n+1}\left(\alpha + \mu I_{\theta}\right)^n; \nonumber \\
\displaystyle\frac{d s}{d\theta} & = & \gamma; \nonumber \\
\displaystyle\frac{d\gamma}{d\theta} & = & -s - \sum_{n=2}^m \mu J_n R_{\oplus}^n \frac{\partial P_n (s)}{\partial s} p_{\lambda}^2 I_{\theta}^{n+3}\left(\alpha + \mu I_{\theta}\right)^{n-1}; \nonumber \\
\displaystyle\frac{d I_{\theta}}{d\theta} & = & \sum_{n=2}^m \mu J_n R_{\oplus}^n \frac{\partial P_n (s)}{\partial s} \gamma I_{\theta}^{n+2}\left(\alpha + \mu I_{\theta}\right)^{n-1}; \nonumber \\
\displaystyle\frac{d \beta}{d\theta} & = & -\sum_{n=2}^m \mu J_n R_{\oplus}^n \frac{\partial P_n (s)}{\partial s} \frac{1}{p_{\lambda}}s \xi I_{\theta}^{n}\left(\alpha + \mu I_{\theta}\right)^{n-1}; \nonumber \\
\displaystyle\frac{d \xi}{d\theta} & = & \sum_{n=2}^m 2\mu J_n R_{\oplus}^n \frac{\partial P_n (s)}{\partial s} \frac{1}{p_{\lambda}^2}\gamma \xi^2 I_{\theta}^{n-1}\left(\alpha + \mu I_{\theta}\right)^{n-1}; \nonumber \\
\displaystyle\frac{dp_{\lambda}}{d\theta} & = & 0.
\end{eqnarray}
which as it can be seen are also complete polynomial in the variables selected since both Legendre polynomials and their derivatives are always in polynomial form.

Finally, as for the $J_2$ problem, we can use the variable $\eta$ as an alternative manner to disambiguate the longitude of the orbit. In particular, the derivative of $\eta$ is:
\begin{equation}
\displaystyle\frac{d \eta}{d\theta} = -\sum_{n=2}^m \mu J_n R_{\oplus}^n \frac{\partial P_n (s)}{\partial s}\frac{1}{s} \frac{1}{p_{\lambda}}\left(\alpha + \mu I_{\theta}\right)^{n-1} \left(\frac{1}{p_{\lambda}^2}I_{\theta}^{n-2}\xi - I_{\theta}^{n}\left(1-s^2\right)\right).
\end{equation}
As it can be seen in the expression, using $\eta$ makes the formulation non-polynomial for the zonal terms of the gravitational potential that have an odd order. Nevertheless, this property allows us to represent in polynomial form the differential equation of $\eta$ for the $J_2$ problem.


\section{Summary of the transformations and differential equations}

In this section we present a summary of the orbital elements proposed in this manuscript as well of their transformations from spherical coordinates and the differential equations resultant of applying a perturbed zonal dynamic into the system. 

\subsection{Orbital elements}

\begin{eqnarray}
\alpha & = & r\sqrt{\dot{\varphi}^2 + \cos^2(\varphi)\dot{\lambda}^2} -  \displaystyle\frac{\mu}{r^2\sqrt{\dot{\varphi}^2 + \cos^2(\varphi)\dot{\lambda}^2}}; \nonumber \\
p_r & = & \dot{r}; \nonumber \\
s & = & \sin(\varphi); \nonumber \\
\gamma & = & \displaystyle\frac{\dot{\varphi}\cos(\varphi)}{\sqrt{\dot{\varphi}^2 + \cos^2(\varphi)\dot{\lambda}^2}}; \nonumber \\
I_{\theta} & = & \displaystyle\frac{1}{r^2\sqrt{\dot{\varphi}^2 + \cos^2(\varphi)\dot{\lambda}^2}}; \nonumber \\
\beta & = & \lambda - \arcsin\left(\tan(\varphi)\sqrt{\frac{\cos^4(\varphi)\dot{\lambda}^2}{\dot{\varphi}^2 + \cos^2(\varphi)\dot{\lambda}^2 - \cos^4(\varphi)\dot{\lambda}^2}}\right); \nonumber \\
\xi & = & \displaystyle\frac{\cos^4(\varphi)\dot{\lambda}^2}{\dot{\varphi}^2 + \cos^2(\varphi)\dot{\lambda}^2 - \cos^4(\varphi)\dot{\lambda}^2}; \nonumber \\
p_{\lambda} & = & r^2\cos^2(\varphi)\dot{\lambda}.
\end{eqnarray}

\subsection{System of 6 variables in non-polynomial form}

\begin{eqnarray}
\displaystyle\frac{d\alpha}{d\theta} & = & -p_r - \sum_{n=2}^{m}\mu J_nR_{\oplus}^n\displaystyle\frac{\partial P_n(s)}{\partial s}\gamma \frac{\left(1-s^2-\gamma^2\right)^{(n+1)/2}}{p_{\lambda}^{n+1}} \nonumber \\
& & \left(\alpha+\mu \frac{\sqrt{1-s^2-\gamma^2}}{p_{\lambda}}\right)^{n-1}\left(\alpha + 2\mu \frac{\sqrt{1-s^2-\gamma^2}}{p_{\lambda}}\right); \nonumber \\
\displaystyle\frac{dp_r}{d\theta} & = & \alpha + \sum_{n=2}^m \left(n+1\right)\mu J_n R_{\oplus}^nP_n(s)\frac{\left(1-s^2-\gamma^2\right)^{(n+1)/2}}{p_{\lambda}^{n+1}} \nonumber \\
& & \left(\alpha + \mu \frac{\sqrt{1-s^2-\gamma^2}}{p_{\lambda}}\right)^n; \nonumber \\
\displaystyle\frac{d s}{d\theta} & = & \gamma; \nonumber \\
\displaystyle\frac{d\gamma}{d\theta} & = & -s - \sum_{n=2}^m \mu J_n R_{\oplus}^n \frac{\partial P_n (s)}{\partial s} \frac{\left(1-s^2-\gamma^2\right)^{(n+3)/2}}{p_{\lambda}^{n+1}} \nonumber \\
& & \left(\alpha + \mu \frac{\sqrt{1-s^2-\gamma^2}}{p_{\lambda}}\right)^{n-1}; \nonumber \\
\displaystyle\frac{d \beta}{d\theta} & = & -\sum_{n=2}^m \mu J_n R_{\oplus}^n \frac{\partial P_n (s)}{\partial s} \frac{s}{s^2+\gamma^2} \frac{\left(1-s^2-\gamma^2\right)^{(n+2)/2}}{p_{\lambda}^{n+1}} \nonumber \\
& & \left(\alpha + \mu \frac{\sqrt{1-s^2-\gamma^2}}{p_{\lambda}}\right)^{n-1}; \nonumber \\
\displaystyle\frac{dp_{\lambda}}{d\theta} & = & 0.
\end{eqnarray}

\subsection{System of 8 variables in polynomial form}

\begin{eqnarray}
\displaystyle\frac{d\alpha}{d\theta} & = & -p_r - \sum_{n=2}^{m}\mu J_nR_{\oplus}^n\displaystyle\frac{\partial P_n(s)}{\partial s}\gamma I_{\theta}^{n+1} \left(\alpha+\mu I_{\theta}\right)^{n-1} \left(\alpha + 2\mu I_{\theta}\right); \nonumber \\
\displaystyle\frac{dp_r}{d\theta} & = & \alpha + \sum_{n=2}^m \left(n+1\right)\mu J_n R_{\oplus}^nP_n(s)I_{\theta}^{n+1}\left(\alpha + \mu I_{\theta}\right)^n; \nonumber \\
\displaystyle\frac{d s}{d\theta} & = & \gamma; \nonumber \\
\displaystyle\frac{d\gamma}{d\theta} & = & -s - \sum_{n=2}^m \mu J_n R_{\oplus}^n \frac{\partial P_n (s)}{\partial s} p_{\lambda}^2 I_{\theta}^{n+3}\left(\alpha + \mu I_{\theta}\right)^{n-1}; \nonumber \\
\displaystyle\frac{d I_{\theta}}{d\theta} & = & \sum_{n=2}^m \mu J_n R_{\oplus}^n \frac{\partial P_n (s)}{\partial s} \gamma I_{\theta}^{n+2}\left(\alpha + \mu I_{\theta}\right)^{n-1}; \nonumber \\
\displaystyle\frac{d \beta}{d\theta} & = & -\sum_{n=2}^m \mu J_n R_{\oplus}^n \frac{\partial P_n (s)}{\partial s} \frac{1}{p_{\lambda}}s \xi I_{\theta}^{n}\left(\alpha + \mu I_{\theta}\right)^{n-1}; \nonumber \\
\displaystyle\frac{d \xi}{d\theta} & = & \sum_{n=2}^m 2\mu J_n R_{\oplus}^n \frac{\partial P_n (s)}{\partial s} \frac{1}{p_{\lambda}^2}\gamma \xi^2 I_{\theta}^{n-1}\left(\alpha + \mu I_{\theta}\right)^{n-1}; \nonumber \\
\displaystyle\frac{dp_{\lambda}}{d\theta} & = & 0.
\end{eqnarray}


\section{Application of Poincar\'e-Lindstedt method to the J2 formulation}

The Poincar\'e-Lindstedt method is a technique to study perturbed periodic systems of ordinary differential equations. The main idea behind this method is to remove the  secular terms arising from the direct application of the perturbation theory from the solution by a proper selection of the frequencies of the problem. That way, the solution remains periodic under the perturbation. 

\subsection{Formulation}

For simplicity of equation manipulation and integration we will use for this section the polynomial formulation in the variables $\{\alpha,p_r,s,\gamma,I_{\theta},\beta,\eta,\xi,p_{\lambda}\}$. In addition, we will focus on the first order solution of the system and show its error performance for typical orbits that can be found in celestial mechanics.

The objective is to find an approximate solution to Eq.~\eqref{eq:j2_poly} defined for the initial conditions $\{\alpha(t=0) = \alpha_0,p_r(t=0)=p_{r0},s(t=0) = s_0,\gamma(t=0) = \gamma_0,I_{\theta}(t=0) = I_{\theta 0},\beta(t=0)=\beta_0,\eta(t=0)=\eta_0,\xi(t=0)=\xi_0,p_{\lambda} = p_{\lambda 0}\}$. To do that we will assume a series solution in the form:
\begin{eqnarray}
\alpha & = & \alpha|_0 + \epsilon \alpha|_1; \nonumber \\
p_r & = & p_r|_0 + \epsilon p_r|_1; \nonumber \\
s & = & s|_0 + \epsilon s|_1; \nonumber \\
\gamma & = & \gamma|_0 + \epsilon \gamma|_1; \nonumber \\
I_{\theta} & = & I_{\theta}|_0 + \epsilon I_{\theta}|_1; \nonumber \\
\beta & = & \beta|_0 + \epsilon \beta|_1; \nonumber \\
\eta & = & \eta|_0 + \epsilon \eta|_1; \nonumber \\
\xi & = & \xi|_0 + \epsilon \xi|_1; \nonumber \\
p_{\lambda} & = & p_{\lambda};
\end{eqnarray}
where $x|_0$ and $x|_1$ are the zero and first order solutions of the variable $x$, and $\epsilon$ is a small parameter, that for the $J_2$ problem has been selected as:
\begin{equation}
\epsilon = \mu J_2 R_{\oplus}^2C_{\theta}^3U,
\end{equation}
where the constant $C_{\theta} = 1/\sqrt{\mu R_{\oplus}}$ is included in order to normalize the perturbing terms, and $U$ is a constant of value 1 whose purpose is to make the expression non-dimensional (units of longitude divided by time). In particular, and for the case of the Earth, the value of the small parameter $\epsilon \approx 1.9833\cdot 10^{-5}$. Also, as part of the normalization of the equations, instead of using the variable $I_{\theta}$, we use the transformed variable $i_{\theta} = I_{\theta}/C_{\theta}$, and thus, we assume a solution for this variable in the form:
\begin{equation}
i_{\theta} = i_{\theta}|_0 + \epsilon i_{\theta}|_1 = \displaystyle\frac{1}{C_{\theta}}I_{\theta}|_0 + \epsilon \displaystyle\frac{1}{C_{\theta}}I_{\theta}|_1 .
\end{equation}
In addition, since in this formulation we have two independent oscillators, we also assume that the variables $\{\alpha,p_r\}$ are periodic with frequency $w_r = w_r|_0 + \epsilon w_r|_1$, while $\{s,\gamma\}$ are periodic with frequency $w_{\varphi} = w_{\varphi}|_0 + \epsilon w_l|_1$, that is:
\begin{eqnarray}
\alpha(\theta = 0) & = & \alpha(\theta = 2\pi w_r); \nonumber \\
p_r(\theta = 0) & = & p_r(\theta = 2\pi w_r); \nonumber \\
s(\theta = 0) & = & s(\theta = 2\pi w_{\varphi}); \nonumber \\
\gamma(\theta = 0) & = & \gamma(\theta = 2\pi w_{\varphi}).
\end{eqnarray}
That way, we can separate the problem into the zero order and first order solution by grouping terms in the powers of $\epsilon$.

\subsubsection{Zero order solution}\label{sec:zero_order}

The zero order solution correspond to this problem of initial conditions:
\begin{equation}
\begin{aligned}[c]
\displaystyle\frac{d\alpha|_0}{d\theta} & = & -p_r|_0; \nonumber \\
\displaystyle\frac{dp_r|_0}{d\theta} & = & \alpha|_0; \nonumber \\
\displaystyle\frac{d s|_0}{d\theta} & = & \gamma|_0; \nonumber \\
\displaystyle\frac{d\gamma|_0}{d\theta} & = & -s|_0; \nonumber \\
\displaystyle\frac{d i_{\theta}|_0}{d\theta} & = & 0; \nonumber \\
\displaystyle\frac{d\beta|_0}{d\theta} & = & 0; \nonumber \\
\displaystyle\frac{d\eta|_0}{d\theta} & = & 0; \nonumber \\
\displaystyle\frac{d\xi|_0}{d\theta} & = & 0;
\end{aligned} \qquad \longrightarrow \qquad \left\{
\begin{aligned}[c]
\alpha|_0 (\theta=0) & = & \alpha_0; \nonumber \\
p_r|_0 (\theta=0) & = & p_{r0}; \nonumber \\
s|_0 (\theta=0) & = & s_0; \nonumber \\
\gamma|_0 (\theta=0) & = & \gamma_0; \nonumber \\
i_{\theta}|_0 (\theta=0) & = & i_{\theta 0}; \nonumber \\
\beta|_0 (\theta=0) & = & \beta_0; \nonumber \\
\eta|_0 (\theta=0) & = & \eta_0; \nonumber \\
\xi|_0 (\theta=0) & = & \xi_0;
\end{aligned} \right.
\end{equation}
whose solution is in fact the non-perturbed solution of the problem:
\begin{equation}
\begin{aligned}[c]
\alpha|_0 & = & \alpha_0\cos(\theta) - p_{r0}\sin(\theta), \nonumber \\
p_r|_0 & = & \alpha_0\sin(\theta) + p_{r0}\cos(\theta), \nonumber \\
s|_0 & = & s_0\cos(\theta) + \gamma_0\sin(\theta), \nonumber \\
\gamma|_0 & = & - s_0\sin(\theta) + \gamma_0\cos(\theta),
\end{aligned} \qquad
\begin{aligned}[c]
i_{\theta}|_0 & = & i_{\theta 0}, \nonumber \\
\beta|_0 & = & \beta_0, \nonumber \\
\eta|_0 & = & \eta_0, \nonumber \\
\xi|_0 & = & \xi_0.
\end{aligned}
\end{equation} 
and thus, $w_{r}|_0 = w_{\varphi}|_0 = 1$.

\subsubsection{First order solution}

The first order solution is defined by this system of differential equations:
\begin{eqnarray}\label{eq:first_order}
\displaystyle\frac{d\alpha|_1}{d\theta} & = & -p_r|_1 + w_r|_1 p_r|_0 + \displaystyle\frac{3}{U}i_{\theta}|_0^3\left(\alpha|_0 + \mu C_{\theta} i_{\theta}|_0\right)\left(\alpha|_0+2\mu C_{\theta} i_{\theta}|_0\right)\gamma|_0 s|_0; \nonumber \\
\displaystyle\frac{dp_r|_1}{d\theta} & = & \alpha|_1 - w_r|_1\alpha|_0+ \frac{3}{2U}i_{\theta}|_0^3\left(\alpha|_0+\mu C_{\theta} i_{\theta}|_0\right)^2\left(3s|_0^2-1\right); \nonumber \\
\displaystyle\frac{d s|_1}{d\theta} & = & \gamma|_1 - w_{\varphi}|_1\gamma|_0; \nonumber \\
\displaystyle\frac{d\gamma|_1}{d\theta} & = & -s|_1 + w_{\varphi}|_1 s|_0 - \displaystyle\frac{3}{U}p_{\lambda}^2C_{\theta}^2i_{\theta}|_0^5\left(\alpha|_0+\mu C_{\theta} i_{\theta}|_0\right)s|_0; \nonumber \\
\displaystyle\frac{di_{\theta}|_1}{d\theta} & = & \displaystyle\frac{3}{U} i_{\theta}|_0^4\left(\alpha|_0 + \mu C_{\theta} i_{\theta}|_0\right)s|_0\gamma|_0; \nonumber \\
\displaystyle\frac{d\beta|_1}{d\theta} & = & - \displaystyle\frac{3}{U}\xi|_0\frac{1}{p_{\lambda}C_{\theta}}i_{\theta}|_0^2\left(\alpha|_0+\mu C_{\theta} i_{\theta}|_0\right)s|_0^2; \nonumber \\
\displaystyle\frac{d\eta|_1}{d\theta} & = & \displaystyle\frac{3}{U}\xi|_0^2\frac{1}{p_{\lambda}C_{\theta}^3}\left(\alpha|_0 + \mu C_{\theta} i_{\theta}|_0\right)\left(\frac{\xi|_0}{p_{\lambda}^2} - C_{\theta}^2 i_{\theta}|_0^2(1-s|_0^2)\right), \nonumber \\
\displaystyle\frac{d\xi|_1}{d\theta} & = & \displaystyle\frac{6}{U}\xi|_0^2\frac{i_{\theta}|_0}{p_{\lambda}^2C_{\theta}^2}\left(\alpha|_0 + \mu C_{\theta} i_{\theta}|_0\right)s|_0\gamma|_0;
\end{eqnarray}
where all the initial conditions of the variables $x|_1$ are set to zero, that is, initially, the orbital perturbation has not altered the solution of the system yet. It is important to note that the zero order solutions are already known from Section~\ref{sec:zero_order} but with the corresponding perturbed frequencies, that is:
\begin{equation}
\begin{aligned}[c]
\alpha|_0 & = & \alpha_0\cos(w_r\theta) - p_{r0}\sin(w_r\theta),  \\
p_r|_0 & = & \alpha_0\sin(w_r\theta) + p_{r0}\cos(w_r\theta),  \\
s|_0 & = & s_0\cos(w_{\varphi}\theta) + \gamma_0\sin(w_{\varphi}\theta),  \\
\gamma|_0 & = & - s_0\sin(w_{\varphi}\theta) + \gamma_0\cos(w_{\varphi}\theta),
\end{aligned} \qquad
\begin{aligned}[c]
i_{\theta}|_0 & = & i_{\theta 0},  \\
\beta|_0 & = & \beta_0,  \\
\eta|_0 & = & \eta_0,  \\
\xi|_0 & = & \xi_0.
\end{aligned}
\end{equation} 

The system of differential equations from Eq.~\eqref{eq:first_order} provides a closed form analytical solution where the variables $\{\alpha, p_r,s,\gamma,\beta,\eta\}$ have secular terms, and thus they are not periodic. Conversely, variables $\beta$ and $\eta$ also show a secular variation, but due to the fact that they are closely related with the right ascension of the ascending node (see Eqs.~\eqref{eq:beta} and~\eqref{eq:eta}), we already know that they are  going to present a secular variation over time. In fact these secular terms cannot vanish no matter the values of $w_r|_1$ and $w_{\varphi}|_1$.

On the other hand, for variables $\alpha$, $p_r$, $s$, and $\gamma$, we can impose that their secular variation vanishes with a proper selection of the perturbing terms in the frequency $w_r|_1$ and $w_{\varphi}|_1$. In particular, these secular terms disappear when these perturbing frequencies are:

\begin{eqnarray}
w_r|_1 & = & \displaystyle\frac{3}{4U}\left(-2\mu C_{\theta}i_{\theta 0}^4 + 3\mu C_{\theta}i_{\theta 0}^4 \gamma_0^2 + 3\mu C_{\theta}i_{\theta 0}^4 s_0^2\right) \nonumber \\
& = & \displaystyle\frac{3\mu}{4UC_{\theta}^3p_{\lambda}^4}\left(1-s_0^2-\gamma_0^2\right)^2\left(-2 + 3 \gamma_0^2 + 3s_0^2\right); \nonumber \\
w_{\varphi}|_1 & = & \displaystyle\frac{3}{2U}p_{\lambda}^2 \mu C_{\theta}^3i_{\theta 0}^6 = \displaystyle\frac{3\mu}{2UC_{\theta}^3p_{\lambda}^4} \left(1-s_0^2-\gamma_0^2\right)^3;
\end{eqnarray}
This means that the two characteristic frequencies of this system for a first order approximation are:
\begin{eqnarray}
w_r & = & 1 + \displaystyle\frac{3\mu\epsilon}{4C_{\theta}^3p_{\lambda}^4}\left(1-s_0^2-\gamma_0^2\right)^2\left(-2 + 3 \gamma_0^2 + 3s_0^2\right)\epsilon  \nonumber \\
& = & 1 + \displaystyle\frac{3\mu^2 J_2 R_{\oplus}^2}{4p_{\lambda}^4}\left(1-s_0^2-\gamma_0^2\right)^2\left(-2 + 3 \gamma_0^2 + 3s_0^2\right); \nonumber \\
w_{\varphi} & = & 1 + \displaystyle\frac{3\mu\epsilon}{2C_{\theta}^3p_{\lambda}^4} \left(1-s_0^2-\gamma_0^2\right)^3\epsilon = 1 + \displaystyle\frac{3\mu^2 J_2 R_{\oplus}^2}{2p_{\lambda}^4} \left(1-s_0^2-\gamma_0^2\right)^3.
\end{eqnarray}

\subsubsection{General first order solution}

In general, the first order solution provided by the Poincar\'e-Lindstedt method contains a large number of terms which makes it unpractical to include the complete solution in this manuscript. However, in order to show the structure of the solution, we have included the expressions of $i_{\theta}$ and $\beta$.

The solution of $i_{\theta}$ is provided by:
\begin{eqnarray}
i_{\theta} & = & i_{\theta 0} -\displaystyle\frac{3\mu J_2 R_{\oplus}^2C_{\theta}^3}{4 w_l (4 w_l^2 - w_r^2)}\Big(-4 \alpha_0 \gamma_0^2 i_{\theta 0}^4 w_l^2 - 4 C_{\theta} \gamma_0^2 i_{\theta 0}^5 \mu w_l^2 + 
4 \alpha_0 i_{\theta 0}^4 s_0^2 w_l^2 \nonumber \\
& + & 4 C_{\theta} i_{\theta 0}^5 \mu s_0^2 w_l^2 - 4 \gamma_0 i_{\theta 0}^4 p_{r0} s_0 w_l w_r + C_{\theta} \gamma_0^2 i_{\theta 0}^5 \mu w_r^2 - 
C_{\theta} i_{\theta 0}^5 \mu s_0^2 w_r^2 \nonumber \\
& + &
4 C_{\theta} \gamma_0^2 i_{\theta 0}^5 \mu w_l^2 \cos(2 w_l \theta) - 
4 C_{\theta} i_{\theta 0}^5 \mu s_0^2 w_l^2 \cos(2 w_l \theta) \nonumber \\
& - &
C_{\theta} \gamma_0^2 i_{\theta 0}^5 \mu w_r^2 \cos(2 w_l \theta) + 
C_{\theta} i_{\theta 0}^5 \mu s_0^2 w_r^2 \cos(2 w_l \theta) \nonumber \\
& + &
4 \alpha_0 \gamma_0^2 i_{\theta 0}^4 w_l^2 \cos(2 w_l \theta) \cos(w_r \theta) -
4 \alpha_0 i_{\theta 0}^4 s_0^2 w_l^2 \cos(2 w_l \theta) \cos(w_r \theta) \nonumber \\
& + &
4 \gamma_0 i_{\theta 0}^4 p_{r0} s_0 w_l w_r \cos(2 w_l \theta) \cos(w_r \theta) -
8 C_{\theta} \gamma_0 i_{\theta 0}^5 \mu s_0 w_l^2 \sin(2 w_l \theta) \nonumber \\
& + &
2 C_{\theta} \gamma_0 i_{\theta 0}^5 \mu s_0 w_r^2 \sin(2 w_l \theta) -
8 \alpha_0 \gamma_0 i_{\theta 0}^4 s_0 w_l^2 \cos(w_r \theta) \sin(2 w_l \theta) \\
& + &
2 \gamma_0^2 i_{\theta 0}^4 p_{r0} w_l w_r \cos(w_r \theta) \sin(2 w_l \theta) -
2 i_{\theta 0}^4 p_{r0} s_0^2 w_l w_r \cos(w_r \theta) \sin(2 w_l \theta) \nonumber \\
& - &
4 \gamma_0^2 i_{\theta 0}^4 p_{r0} w_l^2 \cos(2 w_l \theta) \sin(w_r \theta) +
4 i_{\theta 0}^4 p_{r0} s_0^2 w_l^2 \cos(2 w_l \theta) \sin(w_r \theta) \nonumber \\
& + &
4 \alpha_0 \gamma_0 i_{\theta 0}^4 s_0 w_l w_r \cos(2 w_l \theta) \sin(w_r \theta) +
8 \gamma_0 i_{\theta 0}^4 p_{r0} s_0 w_l^2 \sin(2 w_l \theta) \sin(w_r \theta) \nonumber \\
& + &
2 \alpha_0 \gamma_0^2 i_{\theta 0}^4 w_l w_r \sin(2 w_l \theta) \sin(w_r \theta) -
2 \alpha_0 i_{\theta 0}^4 s_0^2 w_l w_r \sin(2 w_l \theta) \sin(w_r \theta)\Big), \nonumber
\end{eqnarray}
where it can be seen that the expression has no secular terms. Note that this non-secular behavior is also repeated for the variables $\{\alpha,p_r,s,\gamma,\xi\}$. However, for the case of the solution of $\beta$ (and also for $\eta$) we find some secular terms:

\begin{eqnarray} \label{eq:beta_sol}
\beta & = & \beta_0 -\displaystyle\frac{3\mu J_2 R_{\oplus}^2C_{\theta}^3}{4 C_{\theta} w_l w_r (4 w_l^2-w_r^2))} \Big(-8 \xi_0 \gamma_0^2 I_{\lambda} i_{\theta 0}^2 p_{r0} w_l^3 \nonumber \\
& - &
8 \xi_0 I_{\lambda} i_{\theta 0}^2 p_{r0} s_0^2 w_l^3 + 
8 \alpha_0 \xi_0 \gamma_0 I_{\lambda} i_{\theta 0}^2 s_0 w_l^2 w_r \nonumber \\
& + &
8 C_{\theta} \xi_0 \gamma_0 I_{\lambda} i_{\theta 0}^3 \mu s_0 w_l^2 w_r + 
4 \xi_0 I_{\lambda} i_{\theta 0}^2 p_{r0} s_0^2 w_l w_r^2 \nonumber \\
& - &
2 C_{\theta} \xi_0 \gamma_0 I_{\lambda} i_{\theta 0}^3 \mu s_0 w_r^3 + 
8 C_{\theta} \xi_0 \gamma_0^2 I_{\lambda} i_{\theta 0}^3 \mu w_l^3 w_r \theta \nonumber \\
& + &
8 C_{\theta} \xi_0 I_{\lambda} i_{\theta 0}^3 \mu s_0^2 w_l^3 w_r \theta - 
2 C_{\theta} \xi_0 \gamma_0^2 I_{\lambda} i_{\theta 0}^3 \mu w_l w_r^3 \theta \nonumber \\
& - &
2 C_{\theta} \xi_0 I_{\lambda} i_{\theta 0}^3 \mu s_0^2 w_l w_r^3 \theta \nonumber \\
& - &
8 C_{\theta} \xi_0 \gamma_0 I_{\lambda} i_{\theta 0}^3 \mu s_0 w_l^2 w_r \cos(2 w_l \theta) \nonumber \\
& + &
2 C_{\theta} \xi_0 \gamma_0 I_{\lambda} i_{\theta 0}^3 \mu s_0 w_r^3 \cos(2 w_l \theta) \nonumber \\
& + &
8 \xi_0 \gamma_0^2 I_{\lambda} i_{\theta 0}^2 p_{r0} w_l^3 \cos(w_r \theta) \nonumber \\
& + &
8 \xi_0 I_{\lambda} i_{\theta 0}^2 p_{r0} s_0^2 w_l^3 \cos(w_r \theta) \nonumber \\
& - &
2 \xi_0 \gamma_0^2 I_{\lambda} i_{\theta 0}^2 p_{r0} w_l w_r^2 \cos(w_r \theta) \nonumber \\
& - &
2 \xi_0 I_{\lambda} i_{\theta 0}^2 p_{r0} s_0^2 w_l w_r^2 \cos(w_r \theta) \nonumber \\
& - &
8 \alpha_0 \xi_0 \gamma_0 I_{\lambda} i_{\theta 0}^2 s_0 w_l^2 w_r \cos(2 w_l \theta) \cos(w_r \theta) \nonumber \\
& + &
2 \xi_0 \gamma_0^2 I_{\lambda} i_{\theta 0}^2 p_{r0} w_l w_r^2 \cos(2 w_l \theta) \cos(w_r \theta) \nonumber \\
& - &
2 \xi_0 I_{\lambda} i_{\theta 0}^2 p_{r0} s_0^2 w_l w_r^2 \cos(2 w_l \theta) \cos(w_r \theta) \nonumber \\
& - &
4 C_{\theta} \xi_0 \gamma_0^2 I_{\lambda} i_{\theta 0}^3 \mu w_l^2 w_r \sin(2 w_l \theta) \nonumber \\
& + &
4 C_{\theta} \xi_0 I_{\lambda} i_{\theta 0}^3 \mu s_0^2 w_l^2 w_r \sin(2 w_l \theta) \nonumber \\
& + &
C_{\theta} \xi_0 \gamma_0^2 I_{\lambda} i_{\theta 0}^3 \mu w_r^3 \sin(2 w_l \theta) \nonumber \\
& - &
C_{\theta} \xi_0 I_{\lambda} i_{\theta 0}^3 \mu s_0^2 w_r^3 \sin(2 w_l \theta) \nonumber \\
& - &
4 \alpha_0 \xi_0 \gamma_0^2 I_{\lambda} i_{\theta 0}^2 w_l^2 w_r \cos(w_r \theta) \sin(2 w_l \theta) \nonumber \\
& + &
4 \alpha_0 \xi_0 I_{\lambda} i_{\theta 0}^2 s_0^2 w_l^2 w_r \cos(w_r \theta) \sin(2 w_l \theta) \nonumber \\
& - &
4 \xi_0 \gamma_0 I_{\lambda} i_{\theta 0}^2 p_{r0} s_0 w_l w_r^2 \cos(w_r \theta) \sin(2 w_l \theta) \nonumber \\
& + &
8 \alpha_0 \xi_0 \gamma_0^2 I_{\lambda} i_{\theta 0}^2 w_l^3 \sin(w_r \theta) \nonumber \\
& + &
8 \alpha_0 \xi_0 I_{\lambda} i_{\theta 0}^2 s_0^2 w_l^3 \sin(w_r \theta) \nonumber \\
& - &
2 \alpha_0 \xi_0 \gamma_0^2 I_{\lambda} i_{\theta 0}^2 w_l w_r^2 \sin(w_r \theta) \nonumber \\
& - &
2 \alpha_0 \xi_0 I_{\lambda} i_{\theta 0}^2 s_0^2 w_l w_r^2 \sin(w_r \theta) \nonumber \\
& + &
8 \xi_0 \gamma_0 I_{\lambda} i_{\theta 0}^2 p_{r0} s_0 w_l^2 w_r \cos(2 w_l \theta) \sin(w_r \theta) \nonumber \\
& + &
2 \alpha_0 \xi_0 \gamma_0^2 I_{\lambda} i_{\theta 0}^2 w_l w_r^2 \cos(2 w_l \theta) \sin(w_r \theta) \nonumber \\
& - &
2 \alpha_0 \xi_0 I_{\lambda} i_{\theta 0}^2 s_0^2 w_l w_r^2 \cos(2 w_l \theta) \sin(w_r \theta) \nonumber \\
& + &
4 \xi_0 \gamma_0^2 I_{\lambda} i_{\theta 0}^2 p_{r0} w_l^2 w_r \sin(2 w_l \theta) \sin(w_r \theta) \nonumber \\
& - &
4 \xi_0 I_{\lambda} i_{\theta 0}^2 p_{r0} s_0^2 w_l^2 w_r \sin(2 w_l \theta) \sin(w_r \theta) \nonumber \\
& - &
4 \alpha_0 \xi_0 \gamma_0 I_{\lambda} i_{\theta 0}^2 s_0 w_l w_r^2 \sin(2 w_l \theta) \sin(w_r \theta)\Big),
\end{eqnarray}
where we denoted $I_{\lambda} = 1/p_{\lambda}$. In that sense, we have to note that the variable $\beta$ is directly related to the right ascension of the ascending node of the orbit, which is already known to have a secular variation over time. In fact, the secular terms seen in the solution of $\beta$ are related precisely with this drifting of the orbital plane due to $J_2$ perturbation. In particular, if we only consider the secular terms from Eq.~\eqref{eq:beta_sol}, we obtain a secular variation of $\beta$ equal to:
\begin{equation}
    \displaystyle\frac{d\beta}{d\theta} = -\frac{3}{2}\mu^2 J_2 R_{\oplus}^2I_{\theta}^5p_{\lambda},
\end{equation}
which transformed into keplerian elements provides the following expression:
\begin{equation}
    \displaystyle\frac{d\beta}{d\theta} = -\frac{3}{2}J_2 \left(\frac{R_{\oplus}}{a(1-e^2)}\right)^2\cos(i).
\end{equation}
As it can be seen, this result is equivalent to the well know secular variation of the right ascension of the ascending node if we consider an average motion of the system:
\begin{equation}
    \displaystyle\frac{d\Omega}{d t} = -\frac{3}{2}J_2 \left(\frac{R_{\oplus}}{a(1-e^2)}\right)^2\displaystyle\sqrt{\frac{\mu}{a^3}}\cos(i).
\end{equation}

\subsection{Search of periodic orbits}

The formulation presented in this manuscript can be used to search periodic orbits. In particular, if we impose that both characteristic frequencies of the problem ($w_r$ and $w_{\varphi}$) are equal, we will have a dynamic that repeats periodically in the variables $\{\alpha,p_r,s,\gamma,I_{\theta},\xi\}$ at the same frequency $w_{per} = w_r = w_{\varphi}$. That way, the following relation can be obtained:
\begin{equation}
\left(-2 + 3 \gamma_0^2 + 3s_0^2\right) = 2\left(1-s_0^2-\gamma_0^2\right)
\end{equation}
and thus:
\begin{equation}
s_0^2 + \gamma_0^2 = \displaystyle\frac{4}{5}.
\end{equation}
If we relate this result to classical variables:
\begin{equation}
s_0^2 + \gamma_0^2 = 1 - \displaystyle\frac{p_{\lambda}^2}{p_{\theta}^2} = \sin^2(i) = \displaystyle\frac{4}{5},
\end{equation}
whose solution corresponds to the critically inclined orbits at $i = 63.435 \deg$ and $i = 116.565 \deg$. Note also that the relation:
\begin{equation}
\displaystyle\frac{\left(1-s_0^2-\gamma_0^2\right)^2}{p_{\lambda}^4} = \frac{1}{p_{\theta}^4} = 0,
\end{equation}
allows to make both frequencies equal. This case corresponds to a situation where the angular momentum is infinite, that is, the orbiting object is located at an infinite distance orbiting the celestial body. Note also, that these positions correspond to the stable points of the orbit.

\subsection{Examples of application}

The objective now is to show the performance of the solution provided by the Poincar\'e-Lindstedt method to the $J_2$ formulation presented in this work. To that end, we show three examples of orbits around the Earth corresponding to three very different orbit designs: a near circular orbit, a very eccentric elliptic orbit, and an hyperbolic orbit. We present the results of these examples in the following subsections. In all the cases, the analytical solution is compared with a Runge-Kutta numerical scheme of order 4-5 (Dormand-Prince method) with a relative tolerance of $10^{-13}$. 

\subsubsection{Example: near circular orbit}

\begin{figure}[!ht]
	\centering
	{\includegraphics[width = 0.62\textwidth]{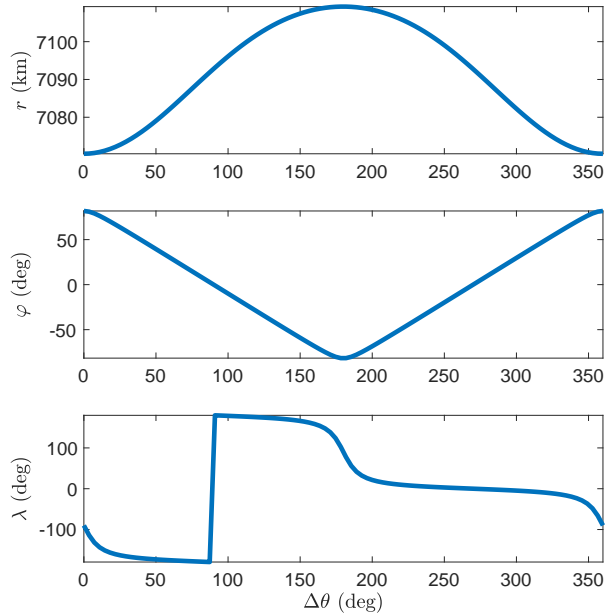}}
	\caption{Evolution of radial distance, latitude and longitude for a frozen sun-synchronous orbit.}
	\label{fig:circ}
\end{figure}

\begin{figure}[!ht]
	\centering
	{\includegraphics[width = 0.62\textwidth]{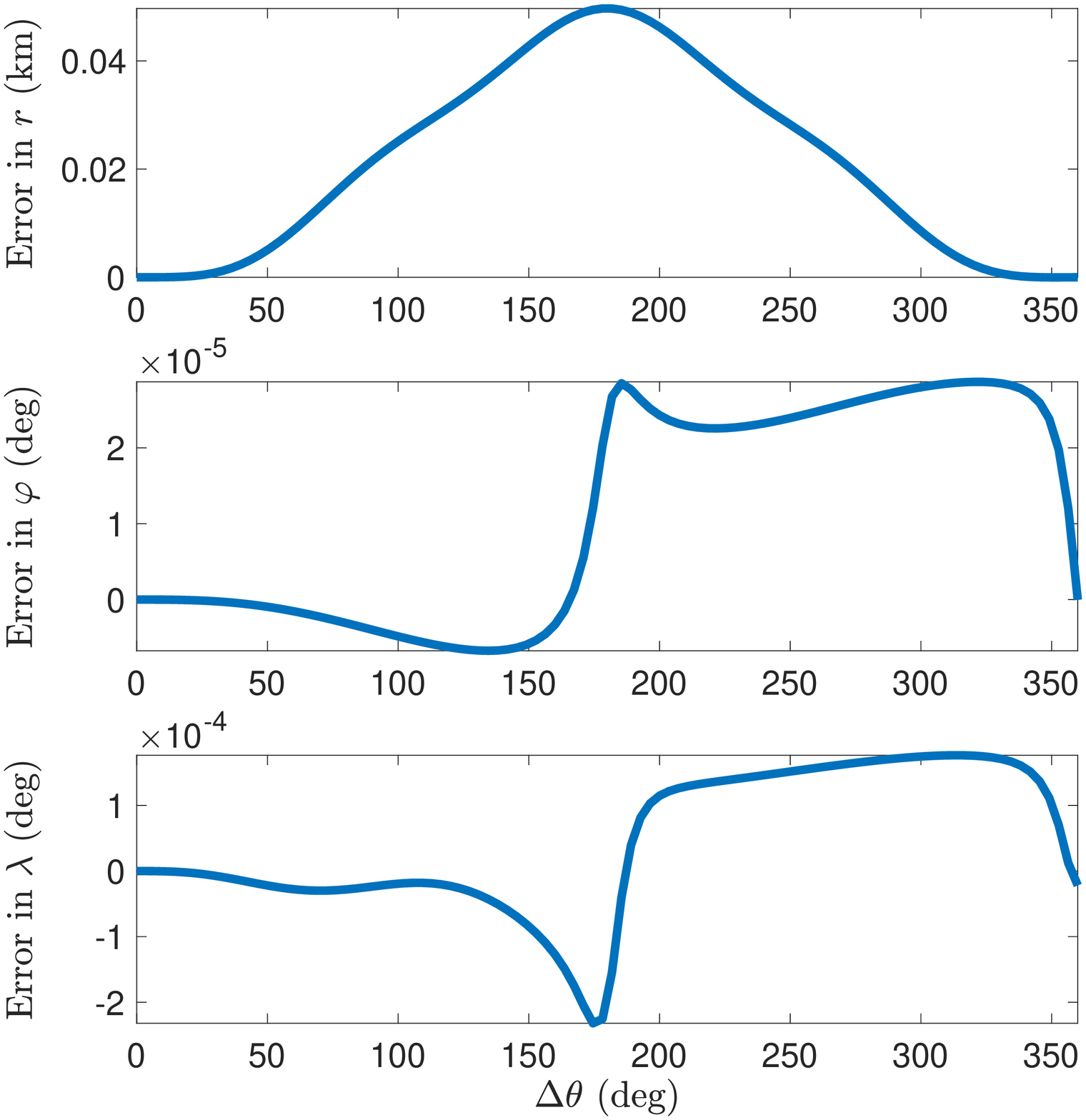}}
	\caption{Evolution of error in radial distance, latitude and longitude for a frozen sun-synchronous orbit.}
	\label{fig:error_circ}
\end{figure}

For the example of near circular orbit, we select a typical sun-synchronous frozen Earth observation orbit with the following initial orbital elements: $a = 7077.722$ km, $e = 0.001043$, $i = 98.186 \deg$, $\omega = 90.0 \deg$, $\Omega = 0.0 \deg$ and $\nu = 0.0 \deg$. Figures~\ref{fig:circ} and~\ref{fig:error_circ} show the evolution of the variables radial distance, latitude and longitude respectively as a function of the time regularization variable $\theta$. These figures also include the error when compared with the numerical solution using a Runge-Kutta scheme. As it can be seen, even for a first order solution of the equations, the precision is remarkable, having less than 50 meters of error in the radial distance, less than $3\cdot10^{-5} \deg$ of error in the latitude of the orbit, and less than $2.5\cdot10^{-4} \deg$ in the determination of the longitude.

\begin{figure}[!ht]
	\centering
	{\includegraphics[width = 0.62\textwidth]{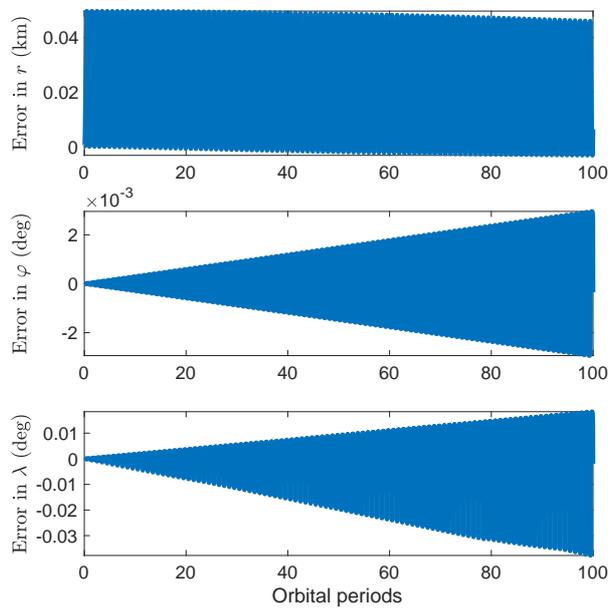}}
	\caption{Long term error evolution in the radial distance, the latitude and the longitude for a frozen sun-synchronous orbit.}
	\label{fig:longterm_circ}
\end{figure}

Additionally, we can study how the error evolves in longer propagation times. To that end, we perform a 100 orbital period propagation using these equations and compare the results with the numerical integrator. Figure~\ref{fig:longterm_circ} shows the result of this comparison. As it can be seen, the error on the radial distance remains bounded while the error in the latitude and the longitude of the orbit increases with time. This effect is due to the first order approximation performed in the frequency of the solution which produces a phasing between the analytical solution and the real solution that increases with time. This effect was observed for all the solutions studied for this work.

\subsubsection{Example: Molniya orbit}

\begin{figure}[!ht]
	\centering
	{\includegraphics[width = 0.62\textwidth]{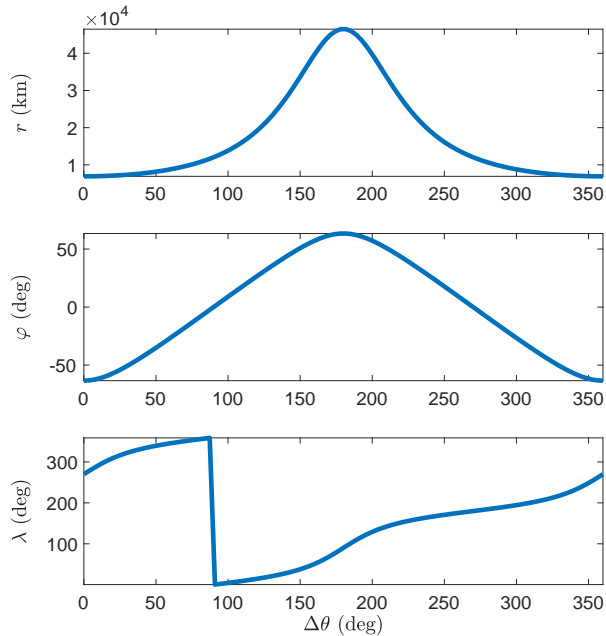}}
	\caption{Evolution of radial distance, latitude and longitude for a Molniya orbit.}
	\label{fig:mol}
\end{figure}

\begin{figure}[!ht]
	\centering
	{\includegraphics[width = 0.62\textwidth]{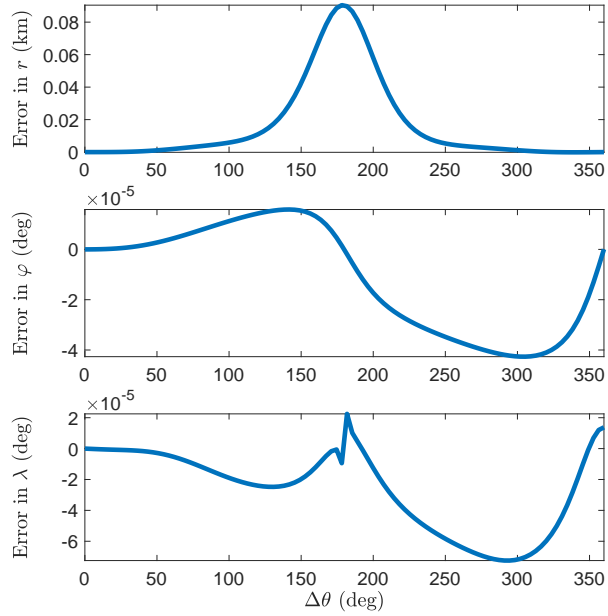}}
	\caption{Evolution of error in radial distance, latitude and longitude for a Molniya orbit.}
	\label{fig:error_mol}
\end{figure}

As an example of this formulation to eccentric orbits, we select a typical Molniya orbit with initial elements: $a = 26600.0$ km, $e = 0.74$, $i = 63.435 \deg$, $\omega = 270.0 \deg$, $\Omega = 0.0 \deg$ and $\nu = 0.0 \deg$. Figures~\ref{fig:mol} and~\ref{fig:error_mol} present the results of applying this formulation to this particular initial conditions. From these figures, it can be seen that, even for very eccentric orbits, the formulation along with the perturbation theory used provides a very good approximation to the problem even for a first order solution. In particular, the order of magnitude of the error in latitude and longitude remains the same as in the near circular case, while the radial distance experiences an increase in its error due to the larger variation of this variable during the dynamic. Therefore, this example shows that this formulation can be applied without any problem even to high eccentric orbits.

\subsubsection{Example: hyperbolic orbit}

\begin{figure}[!ht]
	\centering
	{\includegraphics[width = 0.62\textwidth]{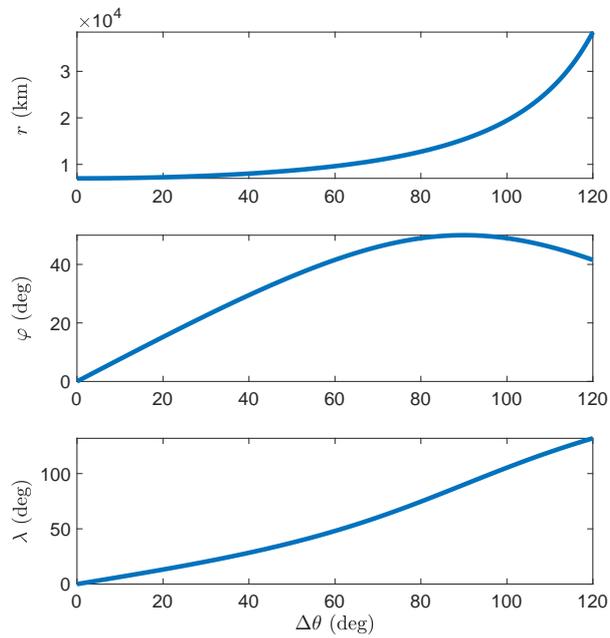}}
	\caption{Evolution of radial distance, latitude and longitude for an hyperbolic orbit.}
	\label{fig:hyp}
\end{figure}

\begin{figure}[!ht]
	\centering
	{\includegraphics[width = 0.62\textwidth]{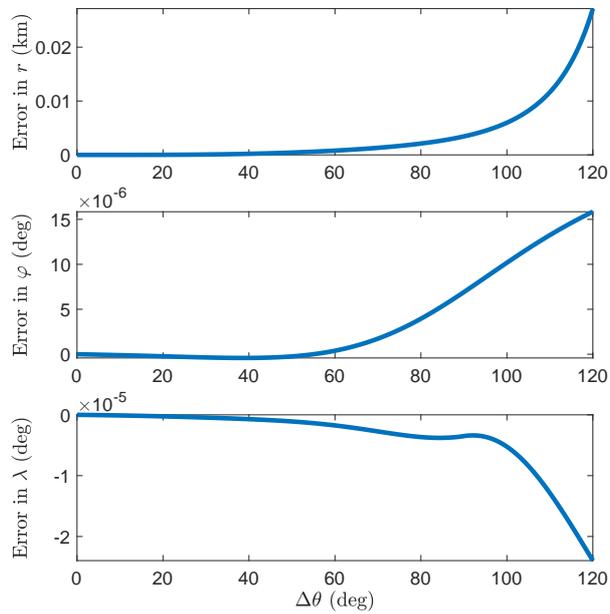}}
	\caption{Evolution of error in radial distance, latitude and longitude for an hyperbolic orbit.}
	\label{fig:error_hyp}
\end{figure}

For the final example, we select an hyperbolic orbit to show that this formulation can also be successfully applied to this kind of orbits. To this end, an orbit with the following initial orbital elements is used: $a = -35000.0$ km, $e = 1.2$, $i = 50.0 \deg$, $\omega = 0.0 \deg$, $\Omega = 0.0 \deg$ and $\nu = 0.0 \deg$. Figures~\ref{fig:hyp}, and~\ref{fig:error_hyp} show the solution for these initial conditions. Note that, since the orbit is hyperbolic, only a section of the orbit was plotted. Nevertheless, the formulation can also generate the imaginary solution when the orbiting object is in the infinite which, to be more precise, generates negative radial distances. As it can be seen from the figures, the precision of the solution is also maintained in this case.

\newpage


\section{Non-dimensional formulation}

Finally, in order to extend the possibilities of application of this set of orbital elements and formulation, we propose in this section the non-dimensional representation to the formulation introduced before. The goal of this new transformation is to normalize all the variables of the problem and to generate a polynomial system of differential equations where all the coefficients are either ones, or multiples of the zonal terms of the gravitational potential from the celestial body in study.

Let $\hat{\alpha}$, $\hat{p}_r$, $\hat{I}_{\theta}$, $\hat{I}_{\lambda}$ be the new set of variables that will substitute $\alpha$, $p_r$, $I_{\theta}$, $I_{\lambda}$ respectively in the equations and that are defined as:
\begin{eqnarray}
&    \hat{\alpha} = \alpha\displaystyle\sqrt{\displaystyle\frac{R_{\oplus}}{\mu}}; \qquad \hat{p}_r = p_r\displaystyle\sqrt{\displaystyle\frac{R_{\oplus}}{\mu}}; & \nonumber \\
& \hat{I}_{\theta} = I_{\theta}\displaystyle\sqrt{\mu R_{\oplus}}; \qquad \hat{I}_{\lambda} = \displaystyle\frac{1}{p_{\lambda}}\displaystyle\sqrt{\mu R_{\oplus}}. &
\end{eqnarray}
Then, if we introduce these transformations in Eq.~\eqref{eq:j2_poly}, the following system of equations is obtained:
\begin{eqnarray}
\displaystyle\frac{d\hat{\alpha}}{d\theta} & = & -\hat{p}_r - 3J_2 \hat{I}_{\theta}^3\left(\hat{\alpha}+\hat{I}_{\theta}\right)\left(\hat{\alpha}+2\hat{I}_{\theta}\right)\gamma s; \nonumber \\
\displaystyle\frac{d\hat{p}_r}{d\theta} & = & \hat{\alpha} + \frac{3}{2} J_2 \hat{I}_{\theta}^3\left(\hat{\alpha}+\hat{I}_{\theta}\right)^2\left(3s^2-1\right); \nonumber \\
\displaystyle\frac{d s}{d\theta} & = & \gamma; \nonumber \\
\displaystyle\frac{d\gamma}{d\theta} & = & -s - 3 J_2 \hat{I}_{\theta}^3\left(1-s^2-\gamma^2\right)\left(\hat{\alpha}+\hat{I}_{\theta}\right)s; \nonumber \\
\displaystyle\frac{d\hat{I}_{\theta}}{d\theta} & = & 3 J_2\hat{I}_{\theta}^4\left(\hat{\alpha} + \hat{I}_{\theta}\right)s\gamma; \nonumber \\
\displaystyle\frac{d\beta}{d\theta} & = & - 3 J_2 \xi\hat{I}_{\lambda}\hat{I}_{\theta}^2\left(\hat{\alpha}+\hat{I}_{\theta}\right)s^2; \nonumber \\
\displaystyle\frac{d\xi}{d\theta} & = & 6 J_2 \xi^2\hat{I}_{\lambda}^2\hat{I}_{\theta}\left(\hat{\alpha} + \hat{I}_{\theta}\right)s\gamma; \nonumber \\
\displaystyle\frac{d\hat{I}_{\lambda}}{d\theta} & = & 0,
\end{eqnarray}
As it can be seen, all the coefficients of the polynomial are ones for the unperturbed terms, and multiples of $J_2$ for the perturbed terms. This formulation can be useful for its application in different perturbation theories that rely on defining the problem based on a perturbation in a small parameter.

In addition, the same transformation can be performed in the general zonal problem from Eq.~\eqref{eq:complete_poly} to obtain:

\begin{eqnarray}
\displaystyle\frac{d\hat{\alpha}}{d\theta} & = & -\hat{p}_r - \sum_{n=2}^{m} J_n\displaystyle\frac{\partial P_n(s)}{\partial s}\gamma \hat{I}_{\theta}^{n+1} \left(\hat{\alpha}+\hat{I}_{\theta}\right)^{n-1} \left(\hat{\alpha} + 2\hat{I}_{\theta}\right); \nonumber \\
\displaystyle\frac{d\hat{p}_r}{d\theta} & = & \hat{\alpha} + \sum_{n=2}^m \left(n+1\right) J_n P_n(s)\hat{I}_{\theta}^{n+1}\left(\hat{\alpha} + \hat{I}_{\theta}\right)^n; \nonumber \\
\displaystyle\frac{d s}{d\theta} & = & \gamma; \nonumber \\
\displaystyle\frac{d\gamma}{d\theta} & = & -s - \sum_{n=2}^m J_n  \frac{\partial P_n (s)}{\partial s} \hat{I}_{\theta}^{n+1}\left(1-s^2-\gamma^2\right)\left(\hat{\alpha} + \hat{I}_{\theta}\right)^{n-1}; \nonumber \\
\displaystyle\frac{d \hat{I}_{\theta}}{d\theta} & = & \sum_{n=2}^m J_n \frac{\partial P_n (s)}{\partial s} \gamma \hat{I}_{\theta}^{n+2}\left(\hat{\alpha} + \hat{I}_{\theta}\right)^{n-1}; \nonumber \\
\displaystyle\frac{d \beta}{d\theta} & = & -\sum_{n=2}^m J_n  \frac{\partial P_n (s)}{\partial s} \hat{I}_{\lambda}s \xi \hat{I}_{\theta}^{n}\left(\hat{\alpha} + \hat{I}_{\theta}\right)^{n-1}; \nonumber \\
\displaystyle\frac{d \xi}{d\theta} & = & \sum_{n=2}^m 2 J_n  \frac{\partial P_n (s)}{\partial s} \hat{I}_{\lambda}^2\gamma \xi^2 \hat{I}_{\theta}^{n-1}\left(\hat{\alpha} + \hat{I}_{\theta}\right)^{n-1}; \nonumber \\
\displaystyle\frac{d\hat{I}_{\lambda}}{d\theta} & = & 0.
\end{eqnarray}
which has again the same property as in the case of the $J_2$ perturbation.


\section{Conclusions}

This manuscript presents a new set of non-singular orbital elements that allows the fully representation and study of the perturbation produced by the zonal harmonics of the gravitational potential about an oblate celestial body. This set of elements are derived from spherical coordinates with respect to the primary body, and provide a linear non-perturbed system when a time regularization is performed. In addition, after performing an expansion of the variables of the problem, the system of differential equations resultant becomes completely polynomial, which provides some advantages for the propagation and also for the application of perturbation theory as we have shown in this manuscript.

In addition, this work includes the application and study of the Poincar\'e-Lindstedt method to the formulation proposed for the particular case of the $J_2$ problem. That approach allows to generate an approximated analytical solution of the problem under $J_2$ perturbation. Particularly, for a motion about the Earth, we show that the formulation has very good accuracy even for a first order solution for orbits at any eccentricity, including elliptic, parabolic and hyperbolic orbits. Note also that the proposed formulation requires no averaging nor removal of the parallax, and thus, it is not required to separate long and short term components of the solution.


\section*{Acknowledgements}

The authors wish to acknowledge useful conversations with Prof. Terry Alfriend at Texas A\&M University on the Poincar\'e-Lindstedt perturbation method.


\printbibliography

\end{document}